# Infusion of Blockchain to Establish Trustworthiness in AI Supported Software Evolution: A Systematic Literature Review

Mohammad Naserameri[1*], Juergen Rilling[1]
[1*]Department of Computer Science and Software Engineering, Concordia University, Montreal, QC, Canada.
*Corresponding author(s). E-mail(s): mohammad.naserameri@mail.concordia.ca; juergen.rilling@concordia.ca

## Abstract

**Context:** Blockchain and AI are increasingly explored to enhance trustworthiness in software engineering (SE), particularly in supporting software evolution tasks.

**Method:** We conducted a systematic literature review (SLR) using a predefined protocol with clear eligibility criteria to ensure transparency, reproducibility, and minimized bias, synthesizing research on blockchain-enabled trust in AI-driven SE tools and processes.

**Results:** Most studies focus on integrating AI in SE, with only 31% explicitly addressing trustworthiness. Our review highlights six recent studies exploring blockchain-based approaches to reinforce reliability, transparency, and accountability in AI-assisted SE tasks.

**Conclusion:** Blockchain enhances trust by ensuring data immutability, model transparency, and lifecycle accountability, including federated learning with blockchain consensus and private data verification. However, inconsistent definitions of trust and limited real-world testing remain major challenges. Future work must develop measurable, reproducible trust frameworks to enable reliable, secure, and compliant AI-driven SE ecosystems, including applications involving large language models.

## 1. Introduction

Bitcoin [1], introduced in 2008, pioneered decentralized data management through blockchain, enabling secure, intermediary-free validation. While initially framed as a financial technology, blockchain's potential expanded with Ethereum in 2015 [2], which introduced smart contracts and spurred applications beyond finance. Large Language Models (LLMs), broadly available since 2022 via ChatGPT, are AI systems trained on vast software and textual datasets [3-5]. They support software engineering (SE) tasks such as code recommendation, documentation, and program comprehension. LLM-powered tools like GitHub Copilot exemplify a new paradigm, termed LLMs for software engineering (LLM4SE) [6,7]. Despite their promise, LLM4SE raises critical trustworthiness challenges, including limited traceability, verifiability, and accountability of generated artifacts [8-10]. Motivated by these challenges and the increasing adoption of blockchain in SE, we conduct a systematic literature review on Blockchain for AI-based Software Engineering Tools and Techniques (BAISET) [11,12]. This review, the first of its kind for software evolution, synthesizes 44 primary studies that explore how blockchain can support trustworthy AI-based SE tools. Our analysis highlights current practices, gaps in empirical evaluation, and opportunities for scalable, secure, and auditable AI-assisted software evolution.

Our work makes the following key contributions:

- **Analysis of publication trends:**
  We analyze publication trends related to BAISET, including research domains, publication venues, and research methodologies. Our SLR provides insights and guidance for both practitioners and researchers, enhancing their understanding of how blockchain technology can be leveraged to establish trustworthiness in AI-based SE tools and techniques, thereby potentially influencing decisions regarding the adoption of LLMs for software evolution tasks.
- **Findings and future research directions:**
  We present key findings and offer recommendations for future research on BAISET, with particular emphasis on the use of blockchain to enhance the trustworthiness of AI-based tools and LLMs that support software evolution activities.

The remainder of this paper is organized as follows. Section 2 reviews related surveys and studies on blockchain and AI-based SE tools and techniques. Section 3 describes the research methodology adopted for the SLR. Section 4 presents the findings, structured according to the research questions. Section 5 discusses the key results and outlines directions for future research. Section 6 addresses the limitations of this study, and Section 7 concludes the paper.

## 2. Related Reviews

Our analysis of related reviews (Table 1) identified several secondary studies on LLM4SE or blockchain in software engineering. Hou et al. [5] conducted an SLR of 395 primary studies, examining how LLMs enhance SE processes, including their applications, performance-improvement strategies, and successful use cases. Zheng et al. [15] reviewed 123 studies, analyzing LLM applications across seven SE tasks such as code generation, summarization, translation, vulnerability detection, evaluation, management, and Q&A while assessing performance and effectiveness. Zhang et al. [16] summarized LLM capabilities and adoption factors in SE, whereas Sasaki et al. [17], analyzing 28 studies, proposed a taxonomy of 21 prompt-engineering patterns grouped into five categories. He et al. [18] reviewed 71 studies on LLM-based multi-agent systems in SE, including maintenance tasks. Fan et al. [19] surveyed emerging LLM4SE research areas and adoption challenges but did not follow a systematic review methodology. Additionally, Gormez et al. [20] conducted a tertiary study of seven SLRs, highlighting LLM capabilities and associated challenges.

Importantly, none of these secondary or tertiary studies explicitly examine blockchain's role in enhancing the trustworthiness of LLM4SE tools. In contrast, our SLR fills this gap by focusing on blockchain-enabled trust in AI-supported software evolution and identifies actionable directions for future research. Table 1 highlights the key differences between prior work and our review.

**Table 1.** State-of-the-art Surveys Related to Blockchain & AI-SE Trustworthiness.

| ID | Reference | Year | Scope | SLR | Time frame | Collected papers |
|---|---|---|---|---|---|---|
| 1 | Sarpatwar et al.[21] | 2019 | Blockchain for AI integrity | NO | Not specified | Not specified |
| 2 | Siddika and Zhao. [22] | 2023 | Blockchain ensures AI/ML data integrity | NO | Not specified | Not specified |
| 3 | Li et al. [23] | 2023 | Trustworthy AI framework | NO | Not specified | Not specified |
| 4 | Ramos and Ellul [24] | 2024 | Blockchain ensures AI compliance. | NO | Not specified | Not specified |
| 5 | Zhang et al. [25] | 2024 | Blockchain-based Trustworthy AI (BTAI) | YES | 2018-2022 | 61 |
| 6 | Khati et al. [12] | 2025 | Trust in LLMs for SE. | YES | 2022-2024 | 88 |
| **Our work** | | **2025** | **Blockchain4Trust in AI-to support Software Evolution related tasks** | **YES** | **2018-2024** | **44** |

## 3. Methodology

This SLR investigates how blockchain technology can be used to enhance the trustworthiness of AI and LLM-powered tools supporting software evolution tasks. We adopt the structured and rigorous methodology introduced to software engineering by Kitchenham et al. [13,14], enabling a systematic and in-depth analysis of prior work through clearly defined research questions. Adherence to these guidelines ensures the replicability of the review and minimizes assessment bias [13].

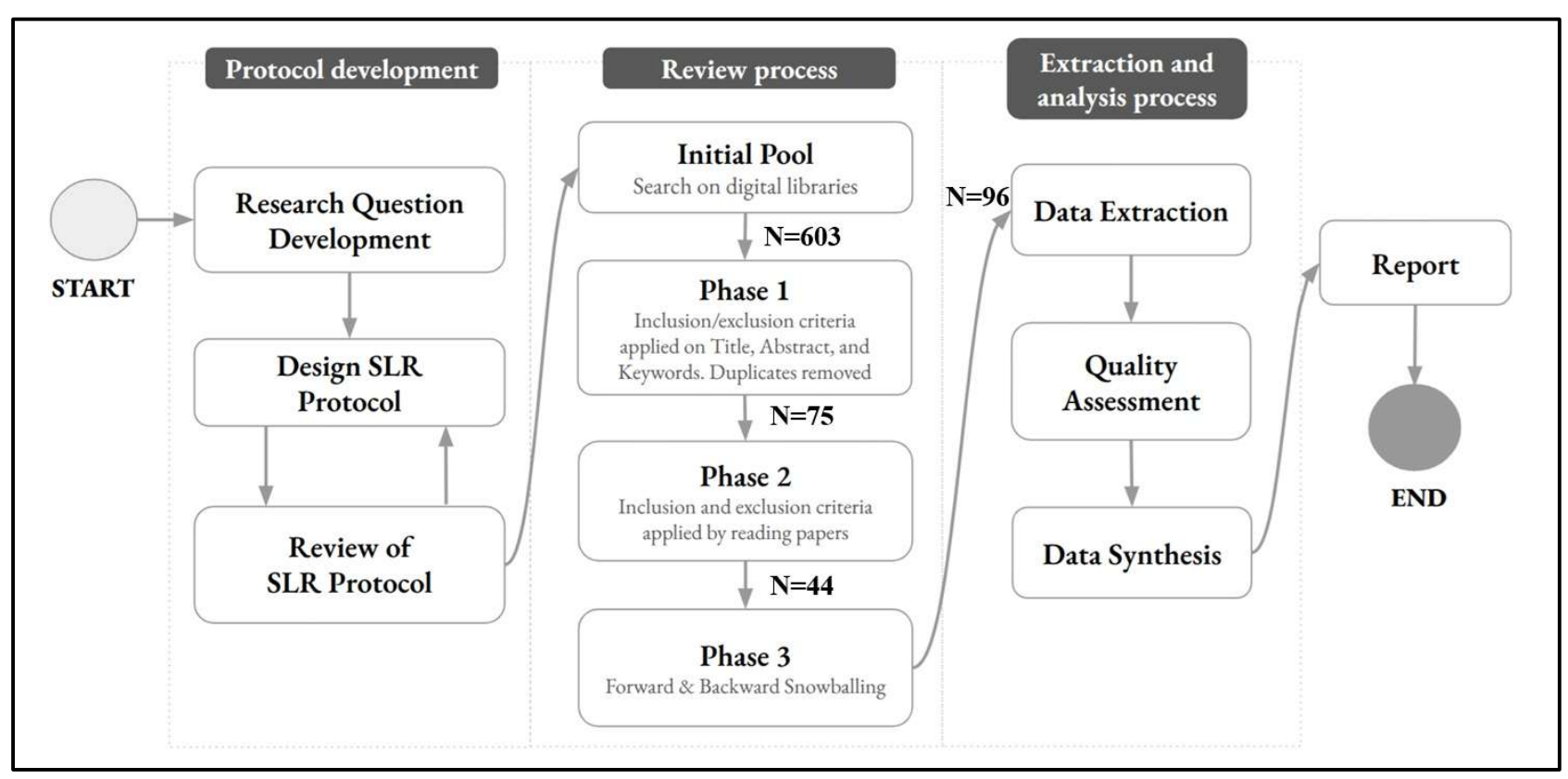

**Figure.1** SLR methodology stages.

Figure 1 presents an overview of the methodology used in our SLR. The first author developed the initial review protocol, which was iteratively refined through discussions with the co-author. The review process was supported by Parsifal[1], a platform for systematic literature reviews, along with spreadsheets to manage snowballing, data extraction, and analysis. The first author executed the database searches, screened and filtered relevant studies, and performed data extraction and analysis, all under the supervision of the co-author. Detailed descriptions of each methodological stage are provided in the following subsections.

## 3.1 Research Questions

Our SLR addresses the following research questions:

**RQ1.** What are the motivations and methodological approaches behind each primary study related to establishing trustworthiness in AI?
The RQ investigates in general how blockchain is used to establish trustworthiness, including in the area of AI. More specifically we look at:

- **RQ1a.** Use of blockchain technology to establish trustworthiness in various domains
  This RQ examines the primary goals, motivations, and methodologies employed by researchers to explore the use of blockchain to establish trustworthiness in various application contexts.
- **RQ1b.** What is the current state of the art in establishing trustworthiness for AI based approaches?
  This RQ examines the primary goals, objectives, motivations, and methodologies employed by researchers in establishing trustworthiness in AI.

**RQ2.** What is the current state of the art in research related to BAISET?
In this RQ, we cover research related to the use of Blockchain to improve Trustworthiness in AI and SE.

- **RQ2.1** How has Blockchain been applied to improve trustworthiness in AI related applications?
- **RQ2.2** How has Blockchain been used to improve trustworthiness in SE related approaches?
- **RQ2.3** Has Blockchain been used to improve the trustworthiness of AI-based SE approaches?

**RQ3**. How is Blockchain used to support LLM4SE in general and more specifically software evolution?
For this research question we focus on the use of Blockchain for LLM4SE and its usage for software evolution related tool support.

**RQ4.** What are limitations and recommendations for future research for the use of BAISET?
For this RQ, we analyze key contributions, limitations and future work discussed by the authors. Based on this, we also suggest future research directions for both practitioners and researchers.

### 3.2. Search Strategy

#### 3.2.1. Data Sources

The research articles included in our SLR were retrieved from widely used and relevant data sources within the SE domain. Specifically, we considered the following sources: the ACM Digital Library, IEEE Xplore, SpringerLink, Elsevier, MDPI, Wiley, Ledger, TU Delft, and Academia. Figure 2 illustrates the distribution of the selected papers across these data sources. To ensure scientific rigor, we excluded non peer-reviewed sources,

[1] Parsifal

such as arXiv. The search was restricted to studies published between 2017 and 2025.

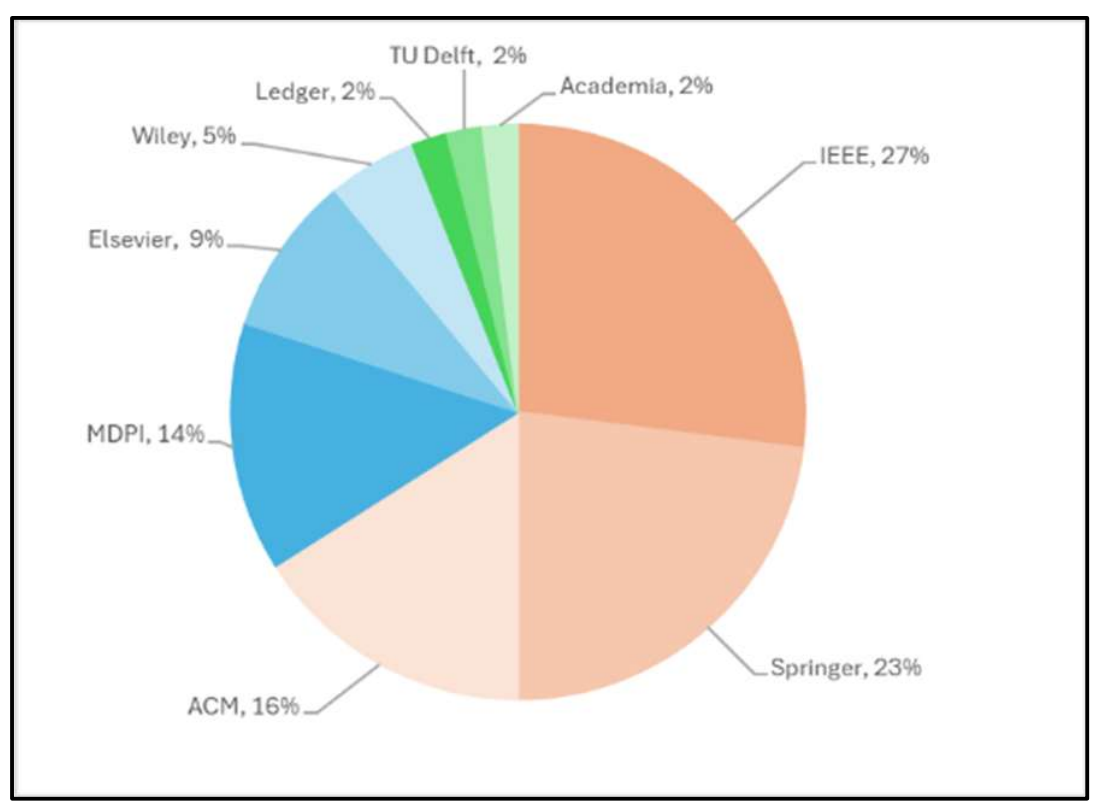

**Figure.2** Distribution of selected papers by database (2017-2025)

### 3.3.2. Search Strings

Following the systematic literature review guidelines by Kitchenham et al. [13,14], we iteratively refined the search strings to accommodate the requirements and constraints of each selected database. Pilot searches were conducted, and the relevance of retrieved studies was evaluated through titles, abstracts, and keywords.

**Table 2.** List of search strings used for each database

| No. | Database | Search string |
|---|---|---|
| 1 | IEEE Xplore | ("Blockchain" AND "AI") OR ("Blockchain" AND "SE") OR ("Blockchain" AND "AI" AND "SE") OR ("AI" AND "trustworthiness") OR ("Blockchain" AND "AI" AND "trustworthiness") filter: Publication Date: 2017-2025; Article Type: Research Article and conference paper |
| 2 | MDPI | ("Blockchain" AND "AI") OR ("Blockchain" AND "SE") OR ("Blockchain" AND "AI" AND "SE") OR ("AI" AND "trustworthiness") OR ("Blockchain" AND "AI" AND "trustworthiness") filter: Publication Date: 2017-2025; Article Type: Research Article and conference paper |
| 3 | Springer Link | ("Blockchain" AND "AI") OR ("Blockchain" AND "SE") OR ("Blockchain" AND "AI" AND "SE") OR ("AI" AND "trustworthiness") OR ("Blockchain" AND "AI" AND "trustworthiness") filter: Publication Date: 2017-2025; Article Type: Research Article and conference paper |
| 4 | Wiley | ("Blockchain" AND "AI") OR ("Blockchain" AND "SE") OR ("Blockchain" AND "AI" AND "SE") OR ("AI" AND "trustworthiness") OR ("Blockchain" AND "AI" AND "trustworthiness") filter: Start Date: 2017-01-01; Article Type: Research Article and conference paper |
| 5 | Elsevier | ("Blockchain" AND "AI") OR ("Blockchain" AND "SE") OR ("Blockchain" AND "AI" AND "SE") OR ("AI" AND "trustworthiness") OR ("Blockchain" AND "AI" AND "trustworthiness") filter: Publication Date: 2017-2025; Article Type: Research Article and conference paper |
| 6 | Ledger | ("Blockchain" AND "AI") OR ("Blockchain" AND "SE") OR ("Blockchain" AND "AI" AND "SE") OR ("AI" AND "trustworthiness") OR ("Blockchain" AND "AI" AND "trustworthiness") filter: Publication Date: 2017-2025; Article Type: Research Article and conference paper |
| 7 | TU Delf | ("Blockchain" AND "AI") OR ("Blockchain" AND "SE") OR ("Blockchain" AND "AI" AND "SE") OR ("AI" AND "trustworthiness") OR ("Blockchain" AND "AI" AND "trustworthiness") filter: Publication Date: 2017-2025; Article Type: Research Article and conference paper |
| 8 | Academia | ("Blockchain" AND "AI") OR ("Blockchain" AND "SE") OR ("Blockchain" AND "AI" AND "SE") OR ("AI" AND "trustworthiness") OR ("Blockchain" AND "AI" AND "trustworthiness") filter: Publication Date: 2017-2025; Article Type: Research Article and conference paper |

This iterative process ensured a balance between sensitivity and precision while consistently capturing studies pertinent to our research questions. For certain databases (e.g., ScienceDirect), search string construction was limited by maximum query length (see the supplementary information package), but all adaptations preserved the original search intent. Table 2 presents the final search strings for each database, supporting transparency and reproducibility of the review process.

### 3.3.3. Inclusion and Exclusion Criteria

During the preparation of this SLR, we defined the inclusion and exclusion criteria (Table 3), which were subsequently applied to guide the paper selection process.

**Table 3.** Inclusion and Exclusion criteria

| ID | Criterion |
|---|---|
| I01 | The paper discusses trustworthiness, traceability, or verifiability in AI-based SE tools or models. |
| I02 | The paper proposes or analyzes the use of blockchain to enhance trust in AI or software systems. |
| I03 | The study addresses at least one of our research questions (RQ1-RQ4). |
| I04 | The paper is written in English. |
| I05 | The full text of the article is accessible. |
| I05 | The paper is not a duplication of others. |
| I07 | The study was published in peer-reviewed journals, conferences, or workshops. |
| I08 | The paper is a primary study (e.g., empirical, case study, SLR, or technical review). |
| I09 | The paper is published between 2017 and 2025. |
| E01 | Short papers with less than four pages. |
| E02 | Papers based solely on opinion without empirical or technical support. |
| E03 | Conference or workshop papers if an extended journal version of the same paper exists. |
| E04 | Non-primary studies (e.g., editorial, blog, tertiary studies). |
| E05 | Papers that are not directly related to trustworthiness, blockchain, or AI-based software tools. |
| E06 | Non-English papers. |

### 3.3.4. Paper Selection

The paper selection process was structured using the following steps:

**Initial Pool:** The search strings were executed across the selected data sources in May 2025, yielding a total of 603 articles.

**Phase 1. Title, Abstract, and Keyword Screening.** In this phase, articles were filtered by examining their titles, abstracts, and keywords, while applying the predefined inclusion and exclusion criteria. Duplicate papers (n = 24) were removed, and additional studies were excluded if their titles, abstracts, and keywords were deemed irrelevant. At the conclusion of this phase, 75 papers were retained for further evaluation.

**Phase 2. Full-Text Screening.** The retained papers were evaluated through a full-text review, applying the inclusion and exclusion criteria consistently. This process resulted in a final set of 44 primary studies. Table 4 reports the number of papers selected from each data source.

**Phase 3 Snowballing.** To complement the automated database search and reduce the risk of omitting relevant studies, manual snowballing was conducted using both backward and forward citation techniques on the 44 primary studies [26, 27], following the recommendations by Wohlin [28]. The snowballing process was organized into four sub-phases:

- **Sub-phase 3.1.** Filtering by title.
- **Sub-phase 3.2.** Filtering by abstract and keywords.
- **Sub-phase 3.3.** Skimming the introduction, methodology, results, and conclusion sections.
- **Sub-phase 3.4.** A final verification step was performed.

At the end of the snowballing process, 52 additional related papers were identified. Table 4 provides a detailed account of the number of papers included at each sub-phase of the snowballing process.

### 3.4. Data Extraction and Synthesis

To address the research questions (RQs), we defined a data extraction form consisting of 36 items, organized into the following five categories:

1. **Bibliographic information**, including paper title, authors, publication year, and publication venue.
2. **Study motivation and methodological rigor**, capturing the approaches used by each primary study to establish trustworthiness (RQ1).
3. **Research on BAISET**, focusing on the integration of blockchain technologies to enhance trustworthiness (RQ2).
4. **Use of blockchain to support LLM4SE**, addressing applications and reported outcomes (RQ3).
5. **Reported limitations and future research directions** related to the use of BAISET (RQ4).

The primary search identified 603 records, which were reduced to 44 studies after applying the inclusion and exclusion criteria across two screening phases (see Table 4). The secondary search, conducted using backward and forward snowballing (Table 5), yielded 3,095 additional records. Following successive snowballing iterations, 20 studies from backward snowballing and 32 studies from forward snowballing were retained. Overall, a total of 96 primary studies were included in the final synthesis.

**Table 4.** Breakdown of papers from the primary search

| | Resource | Initial Pool | Phase 1 | Phase 2 |
|---|---|---|---|---|
| Primary Search | MDPI | 190 | 24 | 6 |
| | IEEE | 120 | 18 | 12 |
| | Springer | 90 | 10 | 10 |
| | ACM | 75 | 7 | 7 |
| | Wiley | 45 | 5 | 2 |
| | ELSEVIER | 35 | 4 | 4 |
| | LEDGER | 15 | 2 | 1 |
| | Academia | 18 | 3 | 1 |
| | TU Delft | 15 | 2 | 1 |
| | **COUNT** | **603** | **75** | **44** |

**Table 5.** Breakdown of papers obtained from the secondary search

| | Resource | Initial Pool | Phase | Phase | Phase | Phase |
|---|---|---|---|---|---|---|
| Secondary Search | | | 3.1 | 3.2 | 3.3 | 3.4 |
| | Backward Snowballing | 3095 | 120 | 45 | 28 | 20 |
| | Forward Snowballing | 0 | 95 | 60 | 42 | 32 |
| | **COUNT** | **3095** | **215** | **105** | **32** | **52** |
| **Combined final paper count (Primary + secondary Search)** | | | | | | **96** |

## 3.5. Paper Assessment

For each selected study, we conducted a quality assessment using a yes-no partial evaluation scheme, following established practices in the software engineering (SE) literature [29, 30]. Eight predefined quality assessment questions (Table 6) guided this evaluation. Consistent with Khalajzadeh and Grundy [27], we also considered the reputation of publication venues using the CORE conference rankings and Scimago Journal Rankings (QA8). Detailed results of the quality assessment are provided in Appendix B.

**Table 6.** Research Paper Evaluation Criteria

| ID | Criterion |
|---|---|
| **QA1** | Is the paper strongly aligned with the proposed SLR? |
| **QA2** | Is the research aim clearly stated? |
| **QA3** | Is there a review of key past work? |
| **QA4** | Is the research methodology clearly defined and aligned with the study's Primary research questions? |
| **QA5** | Does the paper offer adequate details on how the data was collected and analyzed? |
| **QA6** | Are the research findings clearly presented and directly linked to the study's research questions? |
| **QA7** | Does the paper include its limitations, a summary, and suggestions for future research? |
| **QA8** | Is the paper published in a recognized and reputable outlet? |

## 4. Results

In the following subsections, we present the findings of our SLR addressing the four research questions. An online supplementary material package is provided to support further in-depth analysis.

## 4.1 Publication Trends

Across the 44 primary studies included in this SLR (see Fig. 3), journal articles constitute the largest proportion, with 25 studies (56.8%), followed by conference papers (10 studies, 22.7%), book chapters (5 studies, 11.4%), arXiv preprints (3 studies, 6.8%), and magazine articles (1 study, 2.3%). The selected publications span the period from 2017 to 2025.

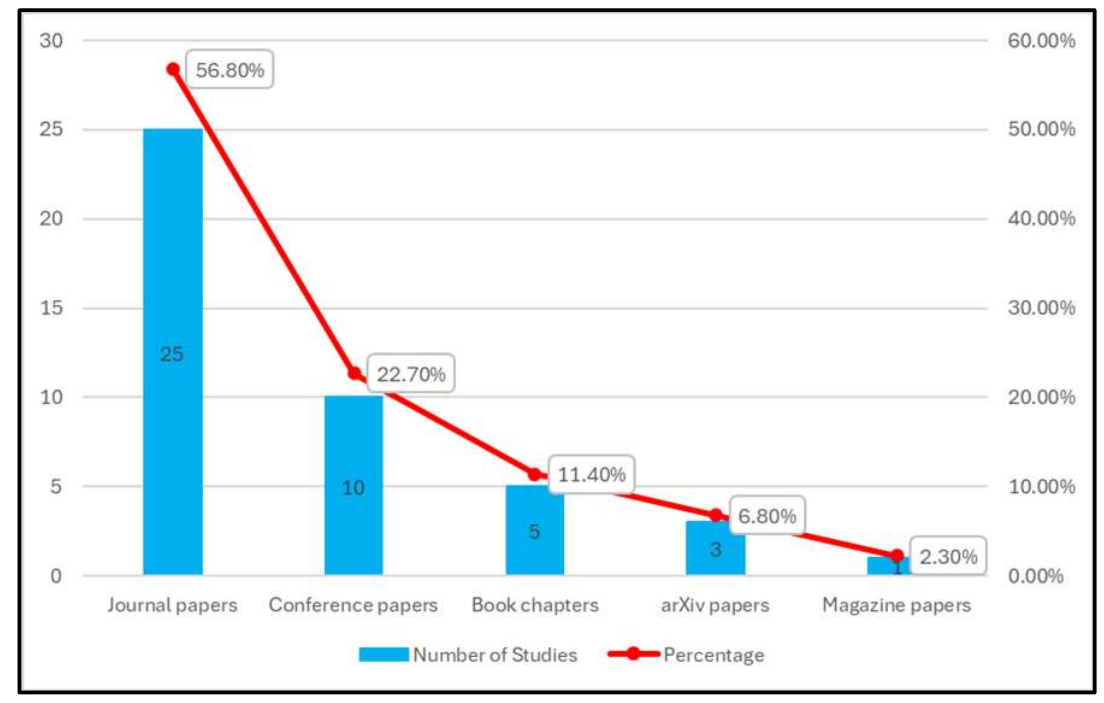

**Figure 3.** Distribution of the 44 primary studies by publication type.

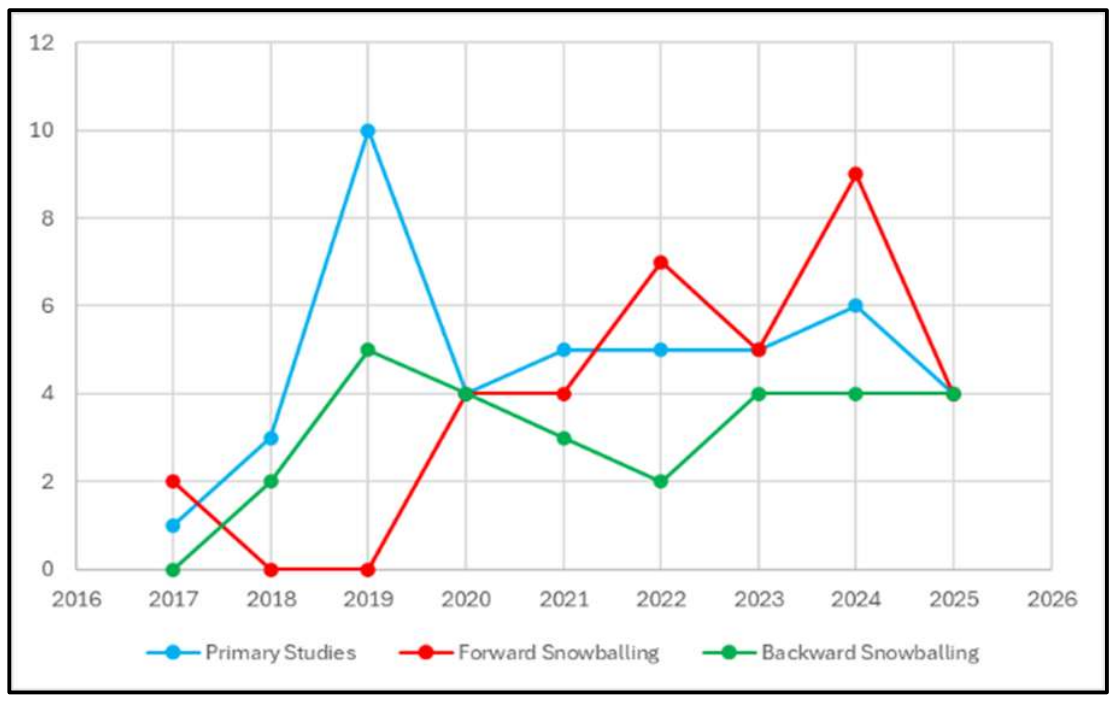

**Figure 4.** Yearly distribution of articles across the three data sources.

As illustrated in Fig. 4, the annual publication volume averages approximately five studies per year. Deviations from this trend occur in 2017, reflecting the early stage and novelty of the research topic, and in 2025, as only publications from the first five months of the year were considered.

We further enriched the SLR dataset by applying forward and backward snowballing as complementary search strategies. Forward snowballing yielded 32 relevant studies (33.3%), while backward snowballing contributed 20 additional studies (20.8%). The primary search accounted for 44 studies (45.9%). Among the surveyed digital libraries, IEEE Xplore, Springer, the ACM Digital Library, and MDPI were the most frequently represented publication venues. Figure 4 provides an overview of the distribution of studies across the different search strategies.The primary dataset exhibits a publication peak in 2019, with 10 studies, indicating early research interest in the synergies between artificial intelligence and blockchain. A more detailed analysis of the forward snowballing dataset reveals a more recent increase in publications during the 2022-2024 period, including nine studies published in 2024, suggesting that recent research increasingly builds upon earlier foundational work. In contrast, the backward snowballing dataset remains relatively stable over the analysis period. Figure 4 also summarizes the annual distribution of included studies across the three data sources (primary search, forward snowballing, and backward snowballing).

Our analysis of the primary studies, based on their titles, abstracts, and keywords, indicates that Blockchain-Oriented Software Engineering (SE) and AI-Enhanced Blockchain Systems are the dominant research domains, accounting for the majority of the included studies (see Fig. 5) and thus playing a central role in the research landscape.

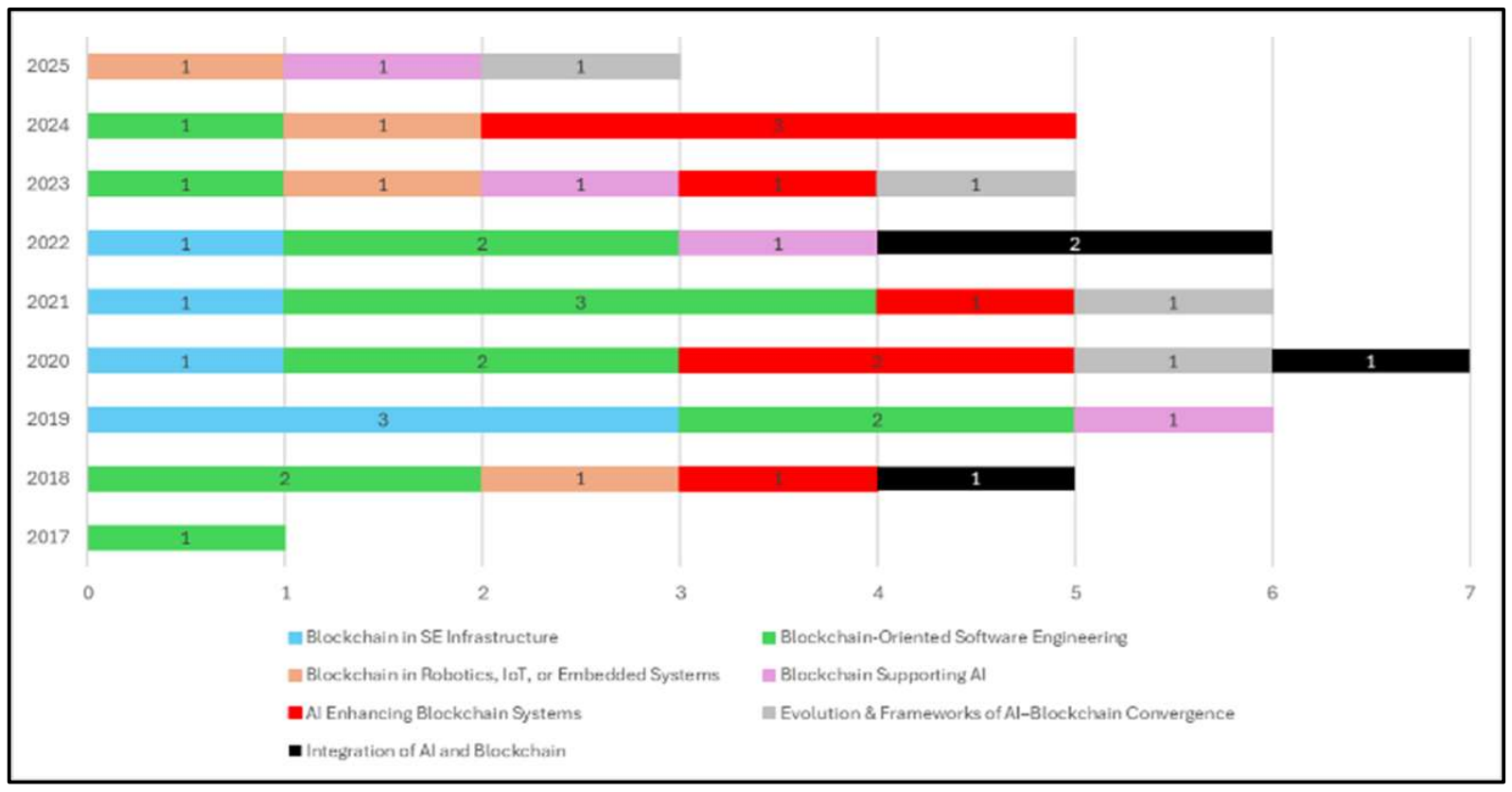

**Figure 5.** Development of research areas over the years.

In contrast, studies addressing emerging domains such as the integration of AI and blockchain, blockchain-supported SE infrastructure, and blockchain applications in robotics, IoT, and embedded systems are beginning to appear. Overall, while the primary studies remain largely concentrated within a limited number of core domains, the findings indicate a gradual diversification of research topics in more recent years.

We further analyzed the primary studies by country of origin and year of publication (see Fig. 6). The majority of the included studies originate from Canadian institutions, accounting for nine studies (20.5%) of the 44 primary studies. Contributions from other countries, including Germany, Italy, and Greece, are comparatively smaller, indicating a more distributed yet still emerging international research interest in this field.

**Figure 6.** Distribution of the selected publications by country and publication year

Table 7 summarizes the distribution of the included studies across recurrent publication venues, focusing on reputable outlets ranked A* or A according to the CORE ranking[2]. Overall, the 44 primary studies are distributed across 27 distinct publication venues. Approximately 61% of the studies are published in either CORE A/A-ranked conferences* or Q1/Q2-ranked journals.

**Table 7.** Publications based on ranked conference (CORE A* or A) or a ranked journal (Q1/Q2)
[ Form Responses sheet, SLR-Data]

| No. | Publication Venue | No. of studies |
|---|---|---|
| 1 | IEEE Journals / Magazines / Conferences | 8 |
| 2 | ACM Journals / Conference | 5 |
| 3 | MDPI Journals | 1 |
| 4 | Elsevier Journals | 3 |
| 5 | Springer Journals | 3 |
| 6 | Other Journals (Wiley, Ledger) | 1 |

With respect to venue type, IEEE publications including journals, magazines, and conference proceedings represent the largest share of the dataset, followed by ACM venues. MDPI journals appear multiple times, most notably Applied Sciences and Information. Among Elsevier journals, several studies are published in high-impact Q1 outlets, including the Journal of Network and Computer Applications, the Journal of Systems and Software, and Expert Systems with Applications. Multiple contributions also appear in Springer journals, such as Philosophy & Technology and Electronic Markets.

Additional studies are published in Wiley journals, including WIREs Data Mining & Knowledge Discovery, as well as in the specialized Ledger journal. For magazine publications, relevant contributions appear in IEEE Computer and IEEE Network. Finally, the dataset includes several Springer book chapters and conference proceedings (e.g., LNCS, LNNS, and Studies in Big Data), which further complement the overall body of evidence.

### 4.2. RQ1. Motivates and Methodological Approaches for Establishing Trustworthiness in AI

[2] A- CORE Conference Ranking Portal: https://portal.core.edu.au/conf-ranks/
B- Journal quartiles based on Scimago Journal Rank (SJR): https://www.scimagojr.com/

**Research Question Context.** RQ1 investigates how **blockchain is used to establish trustworthiness**, including in AI-based systems. Specifically:
**RQ1a** examines the **technologies used to establish trustworthiness** across various application contexts.
**RQ1b** focuses on **how trustworthiness is established in AI approaches**.

Across the 50 studies in the RQ1 dataset, blockchain and AI/ML clearly dominate the technological landscape (Fig. 7). Blockchain is the most widely adopted technology, appearing in 46 primary studies, 8 backward snowballing studies, and 32 forward snowballing studies, underscoring its central role in establishing trustworthiness. AI/ML technologies show a similarly strong presence, with 40 occurrences in primary studies, 12 in backward snowballing, and 30 in forward snowballing. Smart contracts rank third, reported in 16 primary, 6 backward, and 10 forward snowballing studies, highlighting their role in automating and enforcing trust mechanisms. All other technologies are considerably less prevalent. Consensus mechanisms, distributed ledger variants, IoT and agent-based systems, cryptographic techniques, and decentralized storage solutions (e.g., IPFS) appear only sporadically across the evidence base. Overall, Fig. 7 indicates that trustworthiness in BAISET research is predominantly achieved through the combination of blockchain infrastructures, AI-driven analytics, and smart-contract automation, while other technologies play a secondary, supporting role.

### Trust Targets in BAISET Studies

Analysis of the trust targets reported in the RQ1 dataset indicates that the most frequently addressed dimensions are Security (48%), Trust (generic, 48%), and Transparency (42%), followed by Privacy (24%) and Data Integrity (24%). Less frequently reported dimensions include Traceability (14%), Reliability (12%), Accountability (10%), and Explainability, Verifiability, and Auditability (each 6%).

Figure 8 cross-tabulates trust dimensions against the technologies employed. The results show that Blockchain, AI/ML, and Smart Contracts are predominantly used to support Security, Transparency, and Trust, whereas dimensions such as Reliability and Accountability receive comparatively limited coverage. Overall, these findings suggest that BAISET-related studies tend to prioritize core trust properties associated with secure and transparent system behavior, while secondary trust dimensions are addressed less frequently.

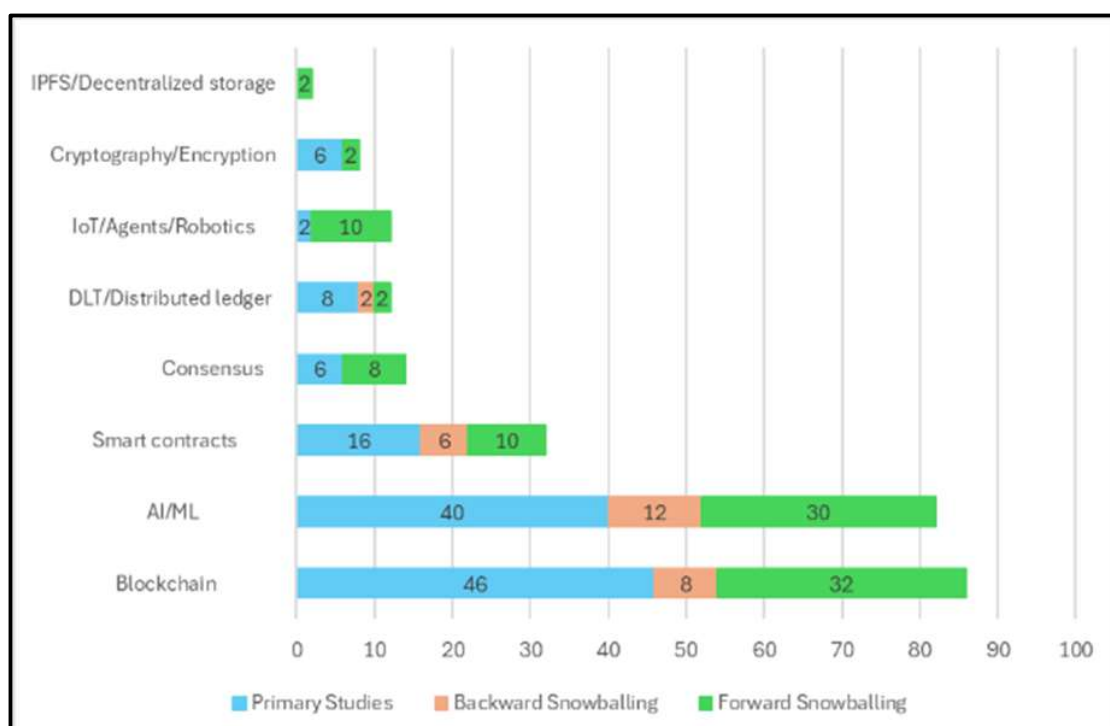

**Figure .7** Technologies Used to Establish Trustworthiness

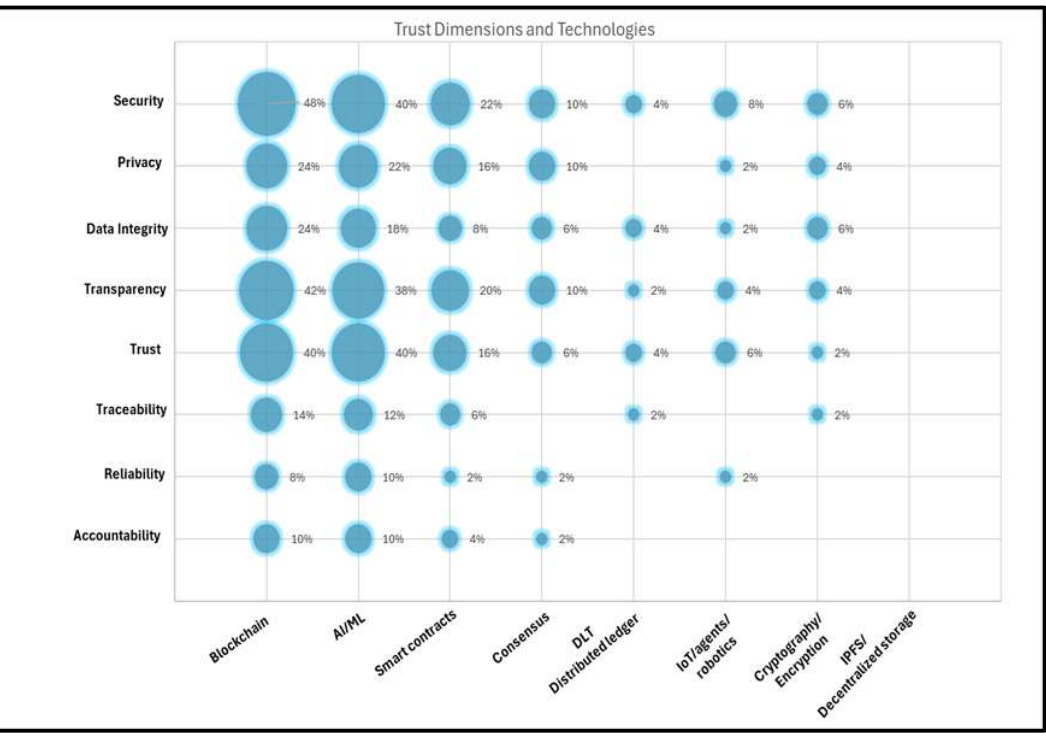

**Figure .8** Co-occurrence of Trust Dimensions and Technologies

### Trustworthiness / Privacy

Our review indicates that most studies prioritize hardening the socio-technical layer notably securing assets and ensuring process visibility over model-centric trust qualities, such as explainability, verifiability, and auditability (each ≤ 6%). Core studies primarily focus on operational safeguards, particularly security and transparency, whereas more recent work increasingly emphasizes trust and reliability. In contrast, backward snowballing studies reflect early-stage concerns, largely centered on foundational trust mechanisms. In practice, blockchain technologies are predominantly used to secure data and enhance system observability. Although privacy and algorithmic accountability are acknowledged, they remain comparatively underdeveloped, highlighting opportunities for privacy-preserving ledger designs, verifiable machine learning, and stronger governance mechanisms.

Across the 50 analyzed studies, trustworthiness is most frequently reported in relation to Security (24 studies, 48%), Trust (generic) (24 studies, 48%), and Transparency (21 studies, 42%), followed by Privacy (12 studies, 24%) and Data Integrity (12 studies, 24%). Less frequently addressed dimensions include Traceability (7 studies, 14%), Reliability (6 studies, 12%), Accountability (5 studies, 10%), and Explainability, Verifiability, and Auditability (3 studies each, 6%). When analyzed by data source, primary studies predominantly cover Security (68%), Transparency (64%), and Trust (44%); forward snowballing studies emphasize Trust (50%) and Reliability (22.2%); and backward snowballing studies focus mainly on Trust (57%) and Transparency (28.6%). Overall, this distribution underscores a strong emphasis on security-, trust-, and transparency-oriented dimensions of trustworthiness across the corpus.

**Framework/tools**

The prevalence of "no specific tool" aligns with prior reviews and conceptual studies, which tend to emphasize processes, architectures, and policy frameworks over concrete implementations. When tools are identified, they primarily support demonstrator prototypes. Examples include Ethereum for smart-contract logic and auditable workflows, Hyperledger for permissioned organizational environments, IPFS for provenance and traceability, and ad hoc contract-verification utilities for Solidity. The cited research frameworks are mostly project-specific conceptual scaffolds rather than reusable, standardized toolkits, limiting replicability and cross-study comparability. This gap highlights an opportunity for the development of standardized, open toolchains and benchmarks in future research.

Among the 50 RQ1a studies, 28 studies (56%) report no specific tool. Named platforms and tools appear only sporadically, including Ethereum (3 studies, 6%), Hyperledger (2 studies, 4%), IPFS (1 study, 2%), contract analysis/verification tools (1 study, 2%), and named research frameworks (3 studies, 6%). When examined by evidence source, "no specific tool" is reported by primary studies (68%), backward snowballing studies (57%), and forward snowballing studies (39%), indicating that forward snowballing studies tend to be more implementation-oriented.

**Type of Evaluation**

The extracted papers predominantly collect evidence through surveys. However, more recent studies, including those identified via forward snowballing, increasingly focus on empirical artifacts and evaluations. Comparative evaluations such as controlled experiments, field studies, or reproducible benchmarks remain relatively rare, underscoring the ongoing need to advance assessment practices beyond narrative surveys toward rigorous, repeatable empirical evaluations of trust mechanisms and their real-world impact. Across the 50 RQ1a studies, evaluation types are distributed as follows Survey (21 studies, 42%), Empirical (16 studies, 32%), Conceptual (11 studies, 22%), and Other (2 studies, 4%).

**Number of studies by evidence source:**

- Primary studies. Survey (12), Conceptual (7), Empirical (5)
- Forward snowballing. Survey (6), Conceptual (2), Empirical (9), Other (2)
- Backward snowballing. Survey (3), Conceptual (2), Empirical (2)

This distribution indicates a gradual shift from survey-based evidence toward more empirical evaluation practices, reflecting a maturing of methodological approaches in the field.

We analyzed the 50 studies in the RQ1a corpus to understand how they design, plan, and execute research and evaluation activities. Activities were grouped into four main categories: Design and Plan, Search and Research, Content Generation, and Analysis and Troubleshooting.

**Design and Plan**

Among the 50 studies, 12 studies (24%) explicitly report design and planning activities Primary studies = 4, Backward snowballing = 4, Forward snowballing = 4. Reported activities include:

- Requirements & architecture definition- 3 studies (Primary = 1, Forward = 2) [M29, F24, F30]
- Methodology/process/workflow design- 5 studies (Primary = 2, Backward = 1, Forward = 2) [M22, M42, B7, F3, F14]
- Governance/policy/risk & threat modeling- 4 studies (Primary = 1, Backward = 3) [M26, B7, B13, B18]

**Search and Research**

30 studies (60%) report evidence-gathering or exploratory research activities - Primary = 15, Backward = 6, Forward = 9. These were further classified as:

- Literature reviews, surveys, SLR/SMS. 18 studies (Primary = 12, Backward = 3, Forward = 3) [M1, M2, M4, M5, M7, M9, M13, M15, M19, M22, M28, M29, B1, B2, B13, F13, F19, F28]
- Application scoping / domain scan (use cases, case studies, domain landscapes). 11 studies (Primary = 4, Backward = 3, Forward = 4) [M9, M13, M29, M32, B1, B4, B8, F9, F24, F27, F29]
- Conceptual analysis / theory building (overviews, perspectives, vision papers). 7 studies (Primary = 3, Backward = 2, Forward = 2) [M9, M13, M26, B1, B18, F25, F32]
- Challenge / gap identification (open issues, future directions). 9 studies (Primary = 7, Backward = 2) [M1, M5, M7, M13, M19, M22, M29, B1, B2]

**Content Generation.**
Among the 50 studies, 38 studies (76%) report producing at least one tangible artifact. Artifacts were grouped into six categories:

- Tools / platforms / prototypes / systems. 9 studies (Primary = 5, Backward = 1, Forward = 3) [M1, M2, M14, M23, M31, B8, F9, F18, F24]
- Frameworks / models / method proposals. 17 studies (Primary = 4, Backward = 5, Forward = 8) [M14, M24, M27, M39, B1, B4, B7, B13, B18, F6, F9, F14, F21, F25, F27, F29, F30]
- Protocols / patterns / smart-contract templates. 17 studies (Primary = 9, Backward = 3, Forward = 5) [M1, M2, M4, M5, M7, M9, M14, M19, M24, B4, B7, B8, F1, F9, F13, F19, F29]
- Datasets / provenance registries / ledgers / logging. 14 studies (Primary = 7, Backward = 1, Forward = 6) [M14, M22, M24, M27, M31, M42, M43, B2, F1, F12, F13, F14, F29, F30]
- Guidelines / taxonomies / checklists / policies. 4 studies (Primary = 1, Backward = 2, Forward = 1) [M39, B2, B13, F25]
- Reference architectures / architectural designs. 3 studies (Primary = 1, Forward = 2) [M30, F24, F30]

**Analysis and Troubleshooting**
37 studies (74%) report some form of diagnostic or evaluation activity (Primary = 19, Backward = 4, Forward = 14). These activities were classified into the following categories:

- Security / threat analysis & detection. 19 studies [M1, M2, M4, M5, M9, M13, M14, M15, M19, M25, M28, M41, M42, B2, F9, F12, F19, F24, F29]
- Traceability / provenance / auditability / accountability. 16 studies [M1, M14, M19, M22, M23, M27, M31, M43, B2, B7, F1, F12, F13, F14, F29, F30]
- Privacy / data-protection assessment. 11 studies [M1, M2, M4, M9, M19, M22, M28, M41, F12, F19, F32]
- Reliability / robustness / fault tolerance. 7 studies [B13, B18, F1, F21, F24, F25, F27]
- Performance / scalability / latency / overhead. 6 studies [M2, M19, M22, F9, F24, F30]
- Verification / validation / consistency / correctness. 5 studies [M14, M31, M43, B18, F25]
- Code / model analysis & understanding (including explainability). 3 studies [M24, F28, F32]

| **RQ1-a.** What are the motivations and methodological approaches behind each primary study related to establishing trustworthiness in AI (Various Domains)? |
|---|
| We categorized 50 studies (25 primary, 7 backward snowballing, and 18 forward snowballing) by trust targets and methodological approaches. Most studies focus on security and threat analysis (38%), followed by traceability, provenance, and accountability (32%) and privacy and data protection (22%). In contrast, reliability and robustness (14%), performance and scalability (12%), and verification and validation (10%) receive substantially less attention.Methodologically, the literature is dominated by artifact-oriented contributions. Studies most often report the development of frameworks, models, or methods (17 studies) and protocols, patterns, or smart-contract templates (17 studies), followed by datasets, provenance registries, or logging mechanisms (14 studies) and tools, platforms, prototypes, or systems (9 studies). By comparison, guidelines, taxonomies, or policies (4 studies) and reference architectures (3 studies) are relatively rare. |

## 4.2 RQ1b - What is the current state of the art in establishing trustworthiness for AI based approaches?

**Technology Adoption**
Across the 40 studies analyzed for RQ1b (Primary = 20, Backward = 5, Forward = 15), AI is the dominant technology, reported in 38 studies (95%), followed by ML in 24 studies (60%) and LLMs in 6 studies (15%). By evidence source, AI remains prevalent across all groups - 19 of 20 Primary studies (95%), all 5 Backward studies (100%), and 14 of 15 Forward studies (93%) employ AI components. ML adoption is more balanced (Primary: 65%, Backward: 60%, Forward: 67%), whereas LLMs show lower overall use but a growing trend, appearing in 2 Primary studies (10%), 1 Backward study (20%), and 3 Forward studies (20%) reflecting increasing interest in integrating LLM-based approaches [M39] (see Fig. 9).

**Tool and Framework Usage**

AI-trustworthiness research is predominantly method- and concept-led rather than tool-led, with authors emphasizing architecture, governance processes, and evaluation themes over reusable engineering artifacts. Notably, the proportion of studies reporting no specific tool/framework decreases in Forward studies (53%), indicating a shift toward implementable solutions, whereas all Backward studies (100%) remain conceptual. This highlights opportunities to develop repeatable toolchains (e.g., auditing, provenance, and verification kits) that operationalize trust across AI lifecycles.

Among the 40 RQ1b studies, 28 (70%) report no specific tool or framework. Named tools appear sporadically: research frameworks in 5 studies (12.5%), other/unspecified tooling in 4 studies (10%), smart-contract/consensus mechanisms in 3 studies (7.5%), and Ethereum in 1 study (2.5%). By evidence source, “no specific tool/framework” is reported in 15 of 20 Primary studies (75%), all 5 Backward studies (100%), and 8 of 15 Forward studies (53%). Categories are not mutually exclusive; a study may report both “no specific tool” and, for example, a consensus mechanism (see Fig. 10).

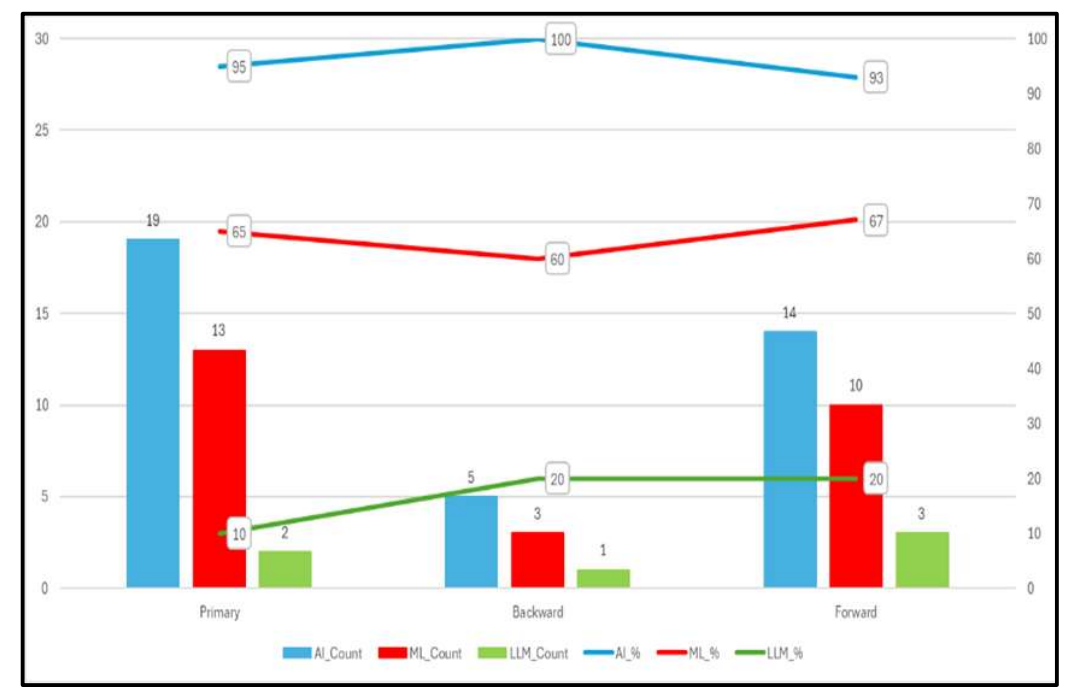

**Fig.9** Technologies used to establish trustworthiness by evidence sources.

**Fig.10** Distribution of Tools and Frameworks Mentioned in Selected Studies.

**Technology (LLM, AI, ML)**

The 40 studies reviewed for RQ1b reveal an AI-centric technology stack - AI is used in 38 studies (95%), ML in 24 studies (60%), and LLMs are explicitly targeted in 6 studies (15%). By evidence source, the 20 Primary studies report AI in 19 studies (95%), ML in 11 studies (55%), and LLMs in 2 studies (10%). Forward snowballing studies (n = 15) show a modest shift toward newer learning stacks - AI in 14 studies (93%), ML in 10 studies (67%), and LLMs in 3 studies (20%). Backward snowballing reflects the foundational baseline - AI in all 5 studies (100%), ML in 3 studies (60%), and LLMs in 1 study (20%). While AI often paired with conventional ML remains the standard approach for establishing trust, LLM-specific work is emerging as an important technology (see Fig.11).

**Types of Trustworthiness (Security, Privacy, Data Integrity)**

Analysis of the 40 studies indicates that trustworthiness is most frequently targeted in the context of:

- Security / threat mitigation 29 studies (72.5%)
- Privacy / data protection 21 studies (52.5%)
- Transparency / traceability / accountability 13 studies (32.5%)
- Data integrity / provenance 8 studies (20%)
- Reliability / robustness 6 studies (15%)
- Explainability / fairness 3 studies (7.5%)

Most papers report on building prototypes or assessing artifacts, yet the primary evidence source remains surveys (20 studies, 50%), with only 9 studies (22.5%) providing empirical evaluations. Artifacts reported include tools/platforms/prototypes/systems in 31 studies (77.5%) and frameworks/models/methods in 17 studies (42.5%).

**Tooling and Frameworks**

Tool support remains limited - 28 studies (70%) report no specific tool or framework. Named tools appear less frequently, including research frameworks (5 studies, 12.5%), consensus/smart-contract mechanisms (3 studies, 7.5%), and Ethereum (1 study, 2.5%).

**Trends Across Evidence Sources**

Within the Primary studies, privacy concerns peak in 13 studies (65%), while security dominates Forward snowballing references, appearing in 13 studies (87%). This shift reflects a movement from foundational concerns toward deployment-stage threat mitigation.

**Key Gaps and Opportunities**

These observations highlight two actionable gaps:

1. **Stronger empirical validation** shared datasets and registries are reported in only 2.5% of studies.
2. **Under-served aspects of human-centered trust** explainability and fairness remain largely neglected despite their importance.

Overall, the current landscape is dominated by AI/ML-based solutions, with emerging adoption of LLMs, strong emphasis on security and privacy, and a need for more systematic empirical validation (Fig.12 ).

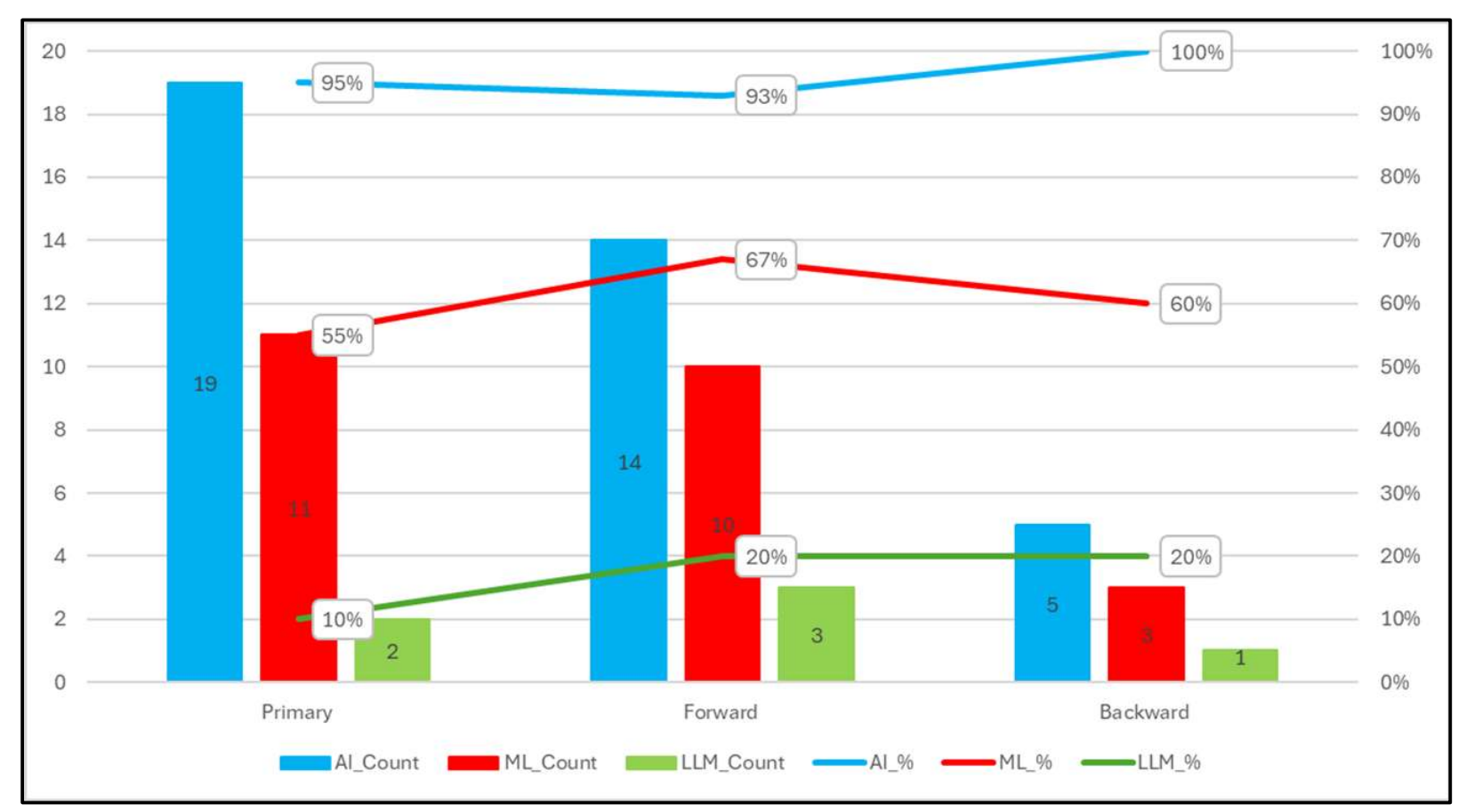

**Fig.11** Technology Usage by Study Group

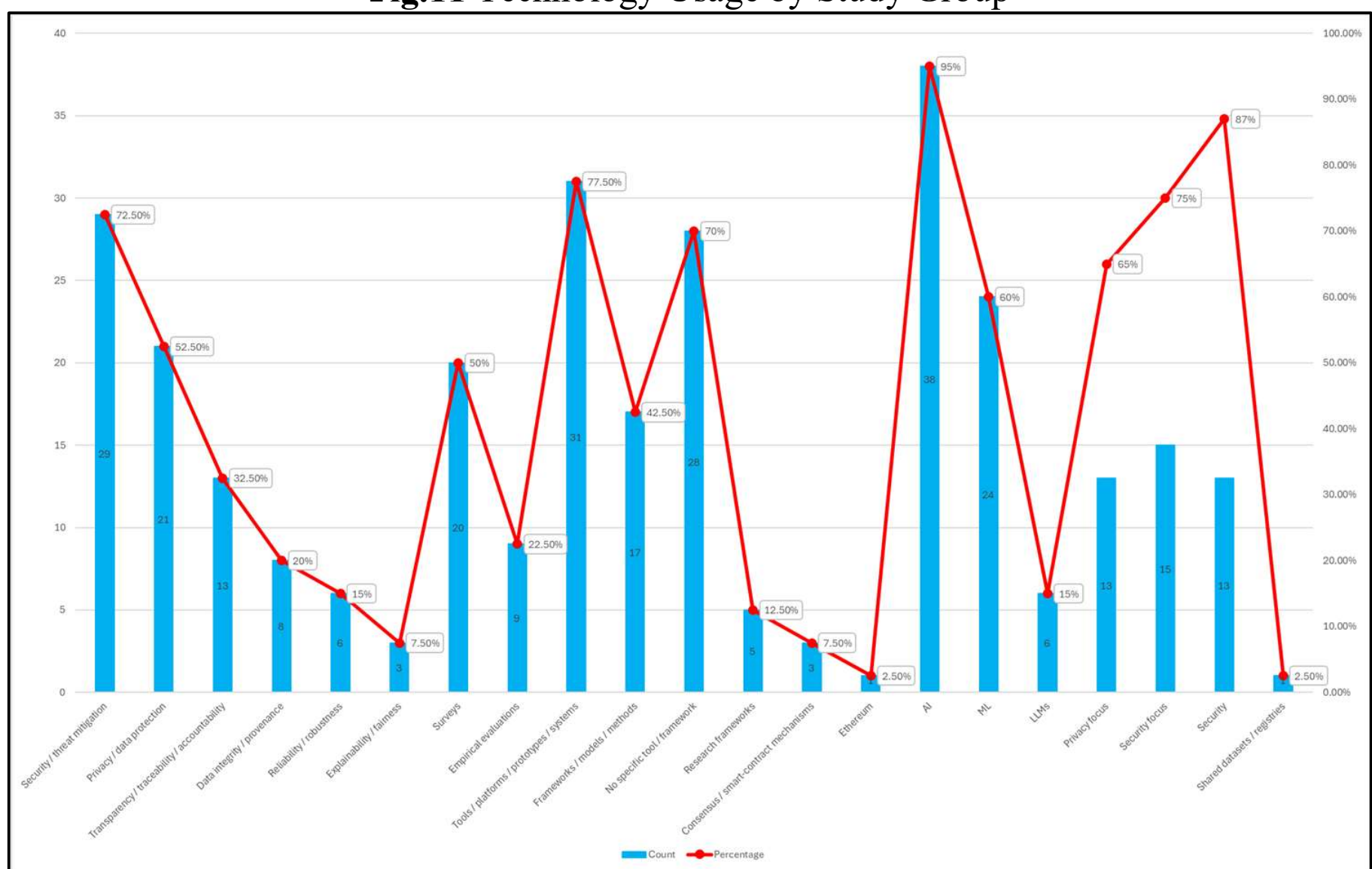

**Fig.12** Distribution of Trustworthiness Targets

**Type of Evaluation**

Evaluations in this domain remain predominantly survey-based, underscoring its largely exploratory maturity. Forward snowballing studies exhibit the highest proportion of empirical evaluations (27%), compared to 20% in both primary and backward datasets, suggesting a recent shift toward experimental assessment. However, large-scale, longitudinal, or deployment-based evaluations are still largely absent. The lack of alternative evaluation types further reveals a gap in controlled experiments, real-world field studies, and benchmark-driven comparisons. Overall, while evaluation practices are becoming incrementally more rigorous, substantial deficiencies in empirical validation persist. Addressing these gaps requires richer experimental designs, real-world case studies, and reproducible benchmarks explicitly aligned with trust objectives.

Within the 40 studies, survey-based research dominates the RQ1b corpus (20 studies, 50%), followed by conceptual papers (11 studies, 27.5%) and empirical studies (9 studies, 22.5%) (see Fig.13).

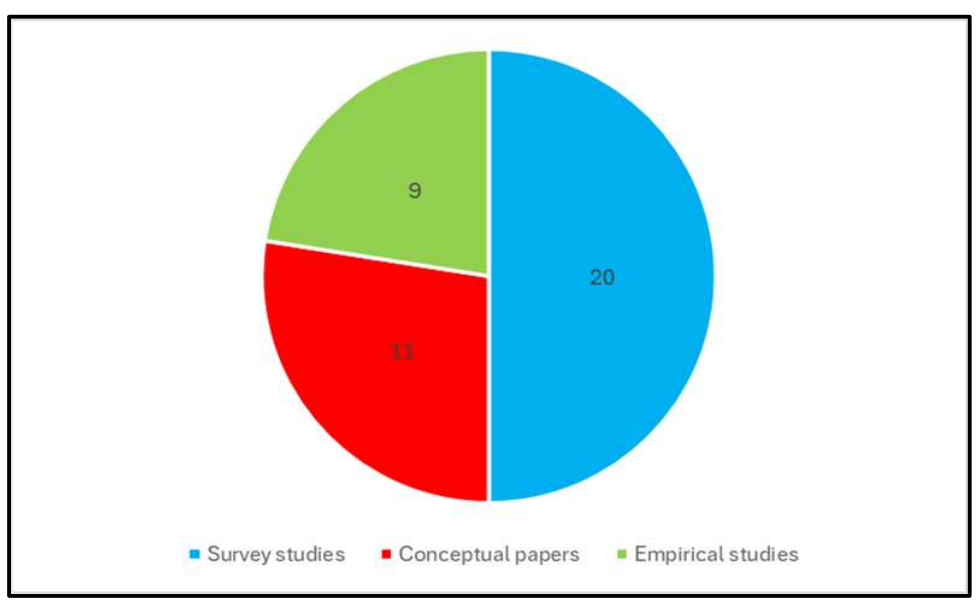

**Fig. 13** Overall Distribution of Evaluation Types

Across the 20 Primary studies, surveys account for 10 studies (50%), followed by conceptual papers with 6 studies (30%) and empirical studies with 4 studies (20%). Among the 15 Forward snowballing studies, surveys appear in 7 studies (46.7%), while empirical and conceptual papers each appear in 4 studies (26.7%). In the 5 Backward snowballing studies, surveys occur in 3 studies (60%), whereas conceptual and empirical studies each occur in 1 study (20%). Within the RQ1b corpus of 40 studies, we identified the following types of evaluation:

**Design and Planning.** Seven studies (17.5%) explicitly report designing or planning evaluation activities (Primary = 2, Backward = 2, Forward = 3; tags are not mutually exclusive). Activities include:

- **Requirements & architecture definition** 3 studies (Primary = 1, Forward = 2) [M24, F21, F25]
- **Methodology / process / workflow design** 2 studies (Backward = 1, Forward = 1) [B18, F25]
- **Governance / policy / risk & threat modeling** 3 studies (Primary = 1, Backward = 2) [M26, B13, B18]

**Searching and researching.** Thirty-one studies (77.5%) report evidence-gathering or exploratory research activities. These are categorized as:

- **Literature reviews / surveys / SLRs** 20 studies (Primary = 11, Backward = 3, Forward = 6) [M1, M2, M4, M7, M28, M29, M33, M34, M40, M41, M44, B1, B2, B13, F6, F12, F13, F19, F25, F32]
- **Application scoping / domain scans** 9 studies (Primary = 4, Backward = 1, Forward = 4) [M1, M2, M34, M41, B1, F19, F24, F27, F28]
- **Conceptual analysis / theory building** 11 studies (Primary = 6, Backward = 2, Forward = 3) [M24, M25, M26, M36, M39, M42, B14, B18, F3, F21, F28]
- **Challenges / gap identification** 6 studies (Primary = 3, Backward = 1, Forward = 2) [M1, M2, M44, B2, F25, F32]

*Note: Studies may appear under multiple tags.*

**Content Generation.** Most studies report producing at least one tangible artifact, which we grouped into six main categories aligned with the corpus:

- **Tools / platforms / prototypes / systems** 13 studies (32.5%) [M2, M24, M30, M31, M39, F1, F8, F9, F21, F24, F27, F29, F30]
- **Framework / model / method proposals** 16 studies (40%) [M24, M27, M30, M31, M36, M39, B18, F3, F8, F9, F12, F21, F24, F25, F27, F30]
- **Protocols / patterns / smart-contract templates** 3 studies (7.5%) [M24, F1, F9]
- **Datasets / provenance registries** 1 study (2.5%) [F30]
- **Guidelines / taxonomies / checklists / policies** 3 studies (7.5%) [M33, B14, F25]
- **Reference architectures / architectural designs** 3 studies (7.5%) [M30, F24, F30]

**Analysis and Troubleshooting.** Across the 40 studies, 32 (80%) report diagnostic or evaluation activities. These can be categorized into corpus-aligned tags:

- **Security / threat analysis and detection** 29 studies (72.5%) [M1, M2, M4, M7, M24, M25, M26, M27, M28, M29, M30, M34, M36, M41, M42, B2, F1, F3, F6, F8, F9, F12, F19, F21, F24, F27, F28, F30, F32]
- **Privacy / data protection assessment** 21 studies (52.5%) [M1, M2, M4, M7, M24, M27, M28, M29, M30, M33, M34, M36, M41, F1, F6, F12, F13, F19, F21, F27, F30]
- **Transparency / traceability / accountability** 13 studies (32.5%) [M1, M7, M26, M27, M28, M31, M41, M42, F1, F6, F8, F12, F21]
- **Reliability / robustness / fault tolerance** 3 studies (7.5%) [M33, F13, F32]
- **Performance / scalability / latency / overhead** 5 studies (12.5%) [M28, M29, M30, F19, F27]
- **Verification / validation / consistency / correctness** 5 studies (12.5%) [M30, M31, F1, F9, F25]
- **Code/model analysis and understanding (including explainability)** 2 studies (5%) [M24, F32]

| **RQ1-b.** What is the current state of the art in establishing trustworthiness for AI-based approaches? |
|---|
| Across 40 studies (20 Primary, 5 Backward, 15 Forward), the field is overwhelmingly **AI-driven** (95%), with **ML** common (60%) and **LLMs still emerging** (15%). Trust efforts focus mainly on **security** (72.5%) and **privacy** (52.5%), while **transparency, provenance, and robustness** receive far less attention. Most work is **artifact-centric**, emphasizing tools and systems (77.5%) and frameworks (42.5%), with limited guidance, datasets, or reference architectures. Evaluation practices indicate **partial maturity** 40% empirical, 35% survey-based, and 22.5% conceptual showing increasing experimentation but persistent gaps in rigorous, large-scale validation. |

**RQ2 What is the current state of the art in research related to BAISET?**

"Blockchain technology provides artificial intelligence with an environment for creating a trusted communication space, which is essential when communication transparency is framed as a trust problem [F3]." In this RQ, we analyze papers exploring how Blockchain is leveraged to enhance trustworthiness in AI, software engineering (SE), or a combination of these areas. As noted, "Artificial intelligence is generating more efficient and secure blockchain protocols as a network, as well as private implementations such as smart contracts [31]." Below, we provide an overview of the analyzed studies.

**RQ2 Tool/Framework.** The RQ2 corpus comprises 49 studies (Primary = 25, Backward = 6, Forward = 18) and reflects an area that is still largely method- and concept-driven rather than tool-oriented. Most papers discuss architectures, governance or process models, and evaluation patterns, without relying on concrete platforms or packages (Fig.14). Where tools are mentioned, they are typically generic enablers such as consensus mechanisms, smart contracts, or occasional Ethereum prototypes rather than full end-to-end trust toolchains. Notably, the Forward set reports "no specific tool" less frequently than the Primary and Backward sets, indicating a gradual shift toward implementable solutions. Opportunities remain for future work, including auditable data/versioning registries and provenance/verification kits that operationalize trust across the AI lifecycle.

Among the 49 studies, 34 (69.4%) do not mention a specific tool or framework, including 19 Primary studies (76%), 5 Backward studies (83.3%), and 10 Forward studies (55.6%). Named research frameworks appear in only 6 studies (12.2%), smart contracts or consensus mechanisms in 4 studies (8.2%), Ethereum explicitly in 1 study (2%), and other/unspecified tooling in 4 studies (8.2%).

**RQ2-Technology**

Across the 49 papers in the RQ2 dataset (Fig.15), AI is the dominant technology, followed by Blockchain. Most studies use AI to generate or verify signals (e.g., predictions), while Blockchain particularly smart contracts and ledgers/consensus mechanisms is employed to make processes tamper-evident, auditable, and traceable. When LLMs are considered, they typically focus on data integrity, auditability, model provenance, or compliant operations (e.g., PoFL-secured training, BC4LLM-style audit trails), reflecting a cautious approach toward generative risks.

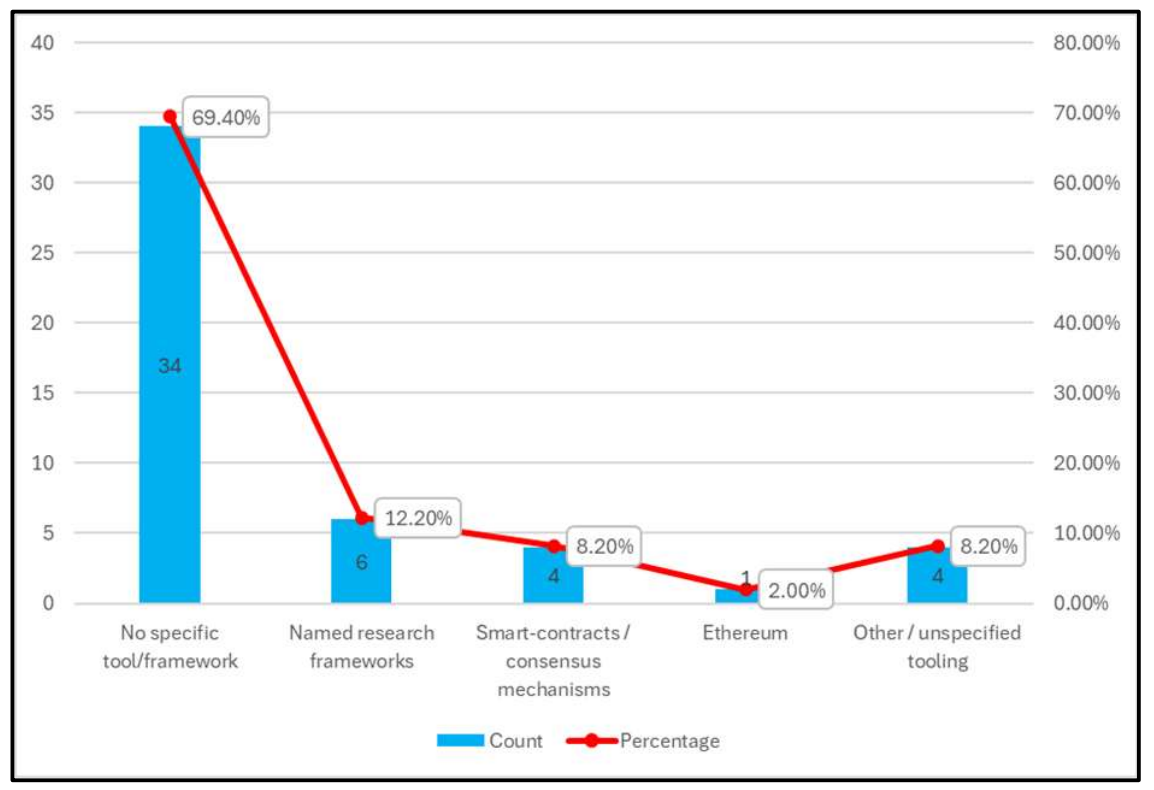

**Fig.14** Distribution of Tools and Frameworks in BAISET Studies

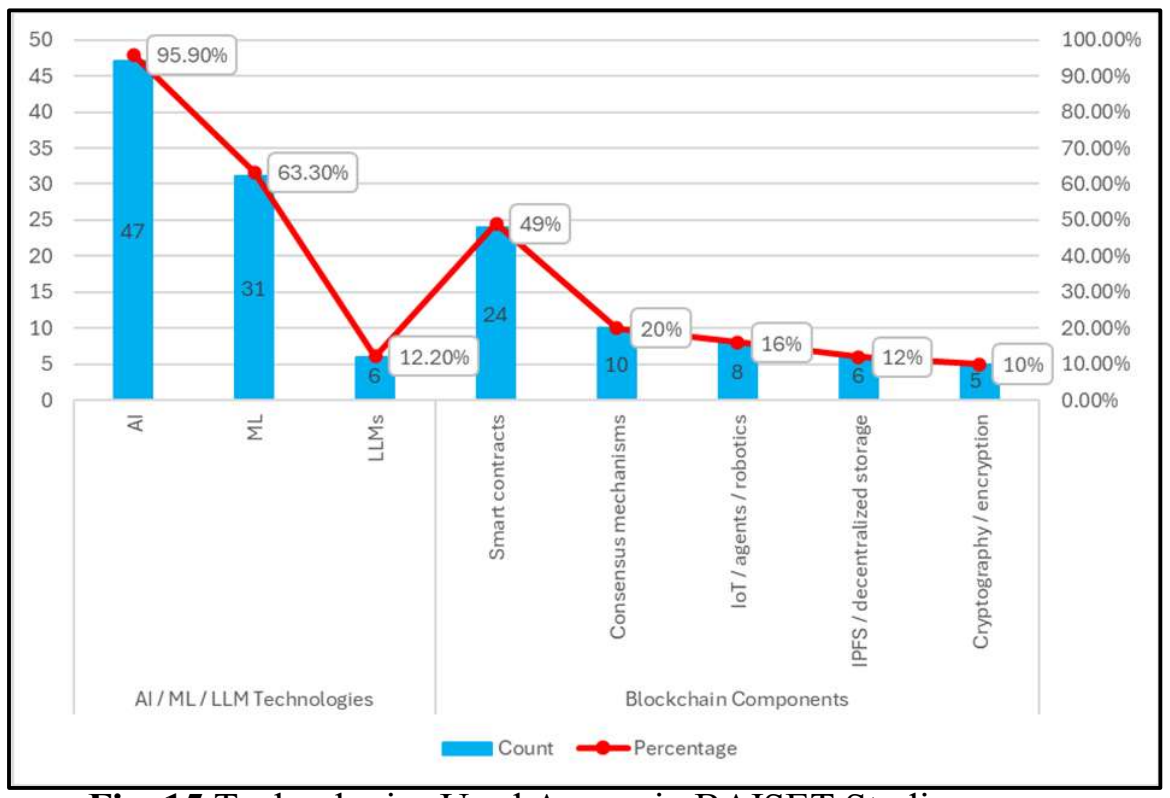

**Fig. 15** Technologies Used Across in BAISET Studies

On the Blockchain side, smart contracts predominate because they operationalize rules (e.g., access, incentives, provenance), with consensus mechanisms and decentralized storage (IPFS) ensuring integrity and reproducibility. A recurring design trade-off noted in these studies is trust versus overhead - stronger on-chain guarantees often conflict with scalability and latency, leading many designs to adopt hybrid on-/off-chain solutions that maintain performance while recording critical trust evidence on-chain.

The technology distribution across the 49 studies is as follows: AI (47 studies, 95.9%), ML (31, 63.3%), and LLMs (6, 12.2%). Common Blockchain building blocks include smart contracts (24, 49%), consensus mechanisms (10, 20%), IoT/agents/robotics (8, 16%), IPFS/decentralized storage (6, 12%), and cryptography/encryption (5, 10%).

**RQ2-Types of Trustworthiness**

Analysis of the 49 studies in the RQ2 corpus reveals that trust is primarily anchored in risk containment for data-centric pipelines. Most papers emphasize confidentiality, integrity, and availability of data flows (e.g., secure training, provenance, auditability), with privacy as a secondary priority (Fig.16). Transparency, traceability, and accountability are treated as enabling layers through mechanisms such as logs, audit trails, and model/data lineage. Reliability, robustness, and human-centric qualities such as explainability and fairness appear far less frequently. Overall, trust is largely operationalized via technical safeguards around data and processes rather than user-oriented assurances, reflecting a security- and privacy-first design approach.

Across the 49 studies, the distribution of trustworthiness types is Security/threat mitigation (36 studies, 73.5%), Privacy/data protection (26, 53.1%), Transparency/traceability/accountability (16, 32.7%), Data integrity/provenance (10, 20.4%), Reliability/robustness/fault tolerance (7, 14.3%), Explainability/XAI (4, 8.2%), and Fairness (4, 8.2%).

Our analysis of evaluation types indicates that the field remains at an early stage of empirical maturity. Fewer than half of the studies report quantitative evaluations such as empirical analyses, prototypes, case studies, or simulations while surveys and conceptual papers make up the majority of the evidence base. This distribution reflects an emerging shift toward implementing blockchain-enabled trust mechanisms for AI (e.g., logging, provenance, and federated-learning accountability), alongside ongoing efforts to consolidate requirements, taxonomies, and governance models. The limited presence of "other/not stated" evaluations further highlights opportunities for more rigorous and reproducible validation, including shared datasets, standardized benchmarks, and real-world deployments.

For the 49 reviewed studies, reported evaluation approaches indicate that the domain is still maturing. Empirical evaluations account for 19 studies (38.8%), surveys (SLR/SMS/scoping) for 17 studies (34.7%), conceptual/theoretical papers for 11 studies (22.4%), and others/not stated for 2 studies (4.1%) (Fig. 17). This distribution suggests growing interest in implementing blockchain-for-AI trust mechanisms (e.g., logging, provenance, federated learning accountability), while many works continue to consolidate requirements, taxonomies, and governance models.

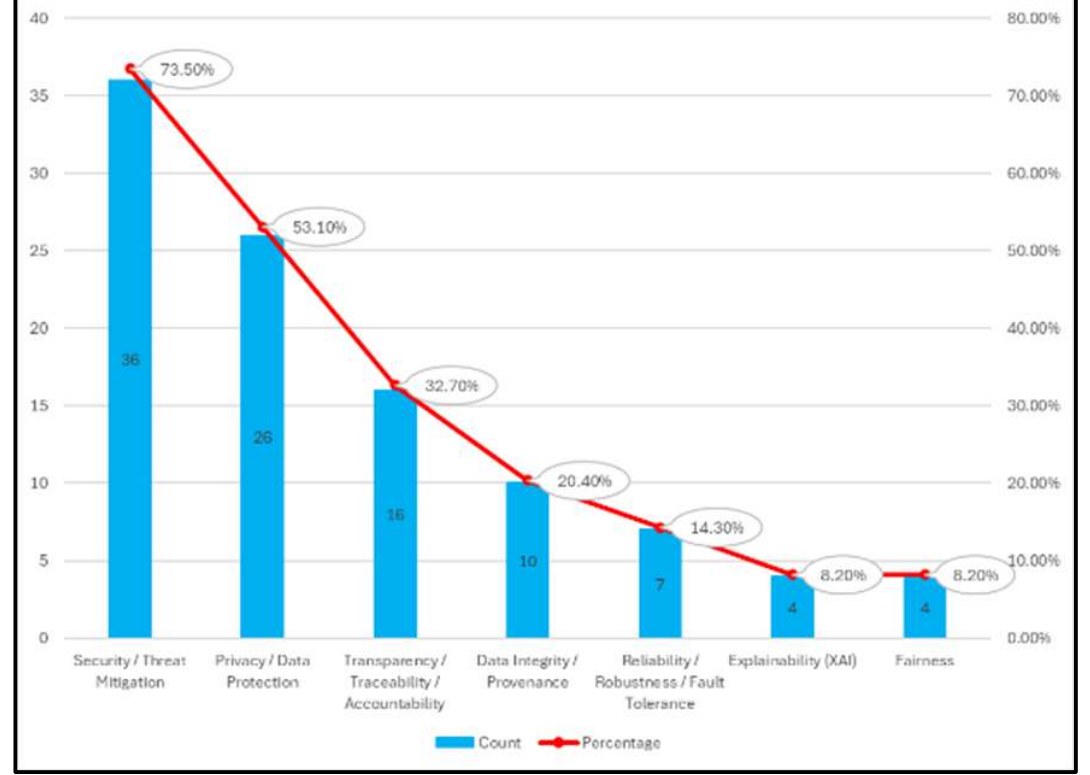

**Fig. 16** Trust Aspects in AI Systems Across BAISET Studies

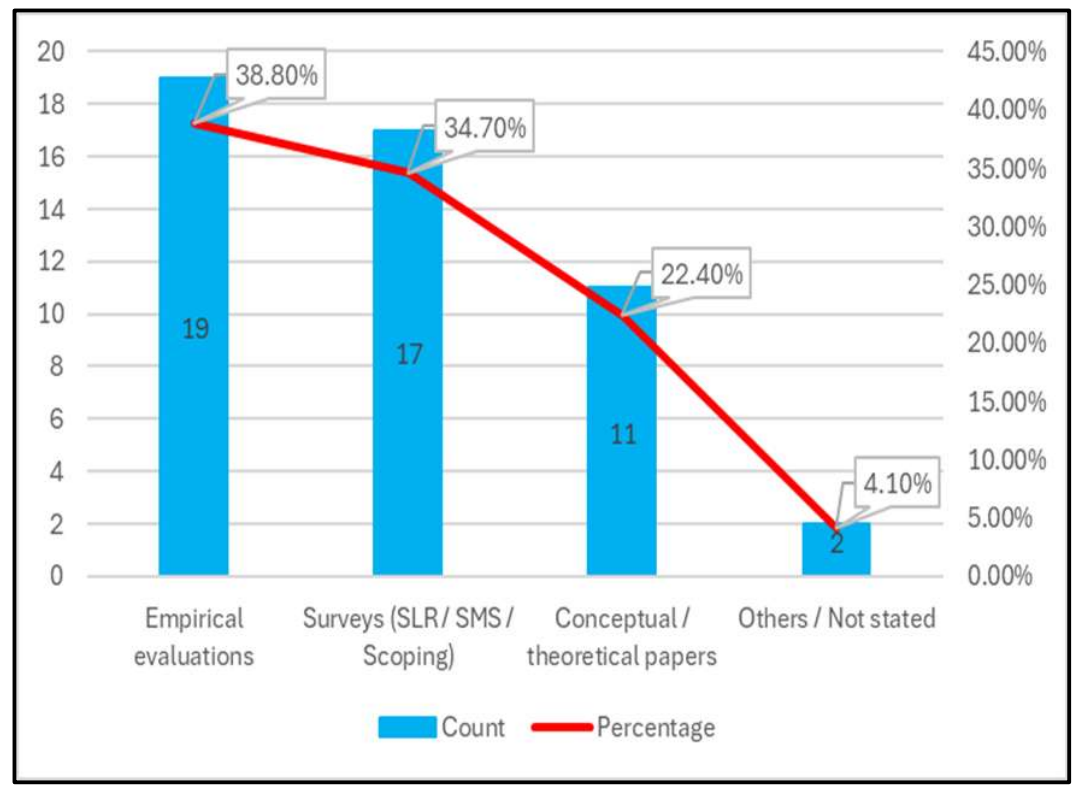

**Fig. 17** Study Types in BAISET Research

**Designing and Planning.** Nine studies (18.4%) emphasize the design and planning of trustworthy AI-Blockchain implementations. These activities are grouped into three areas Requirements and Architecture Definition (4 studies: [M31, M21, B4, F21]), Methodology/Process/Workflow Design (3 studies [M13, M14, B7]), and Governance/Policy/Risk Modeling (2 studies [M26, B8]). These works formalize system blueprints, workflows, and governance models to ensure transparency, auditability, and secure integration.

**Searching and Researching.** Eighteen studies (36.7%) explore the research landscape to map convergence, identify trends, and highlight gaps. This includes Surveys/Systematic Mappings (8 studies, 16.4%), Application Scoping (4 studies, 8.1%), Conceptual Analysis (4 studies, 8.1%), and Challenge Identification (2 studies,4.1%). These studies primarily define the early-stage research space rather than reporting empirical results.

**Content Generation.** Most studies produce at least one tangible output operationalizing AI-Blockchain trust. Tools, Platforms, Prototypes, or Systems appear in 26 studies (54%), Frameworks/Models/Methods in 20 (40%), Protocols/Smart-Contract Templates (5 studies, 10%), and Guidelines/Taxonomies (2 studies, 4%). These artifacts demonstrate functional proof-of-concepts or formal models enforcing trust through transparency and auditability.

**Analysis and Troubleshooting.** Thirty-three studies (67.3%) focus on diagnosing and validating AI-Blockchain trust mechanisms. Key areas include Security/Threat Analysis (28 studies, 57.1%), Privacy/Data Protection (14 studies, 28.5%), Transparency/Traceability/Accountability (16 studies, 33%), Reliability/Robustness (8 studies, 16.3%), and Performance/Scalability/Overhead (6 studies, 12.2%). These efforts reflect the community's focus on securing and validating the operational dependability of integrated systems.

Overall, the frontier between AI and Blockchain is increasingly integrated Blockchain provides an impartial, tamper-evident ledger, while AI leverages this immutable data for intelligent automation, accurate predictions, and trustworthy decision-making. Smart contracts and data pipelines form the primary link between the two technologies. While reviewed studies show early maturity and some verifiable evaluations, significant gaps remain in improving trust assessment standards and addressing challenges such as energy consumption and scalability.

**RQ2.1 How has Blockchain been applied to improve trustworthiness in AI related applications?** In the following, we present a detailed analysis of how Blockchain is applied to enhance the trustworthiness of AI and its related applications. This analysis is motivated by prior findings that "most existing methods for managing the ownership, trading, and access of AI models fall short of providing traceability, transparency, auditability, security, and trust-related features [32]."

**Technology used**

Our review shows that existing research predominantly leverages Blockchain in combination with AI to strengthen trust in data, models, and operational processes [33]. Trust is most commonly operationalized through smart contracts, which enforce auditable, tamper-resistant actions and governance rules [34]. An emerging line of work integrates Blockchain with federated learning to enable privacy-preserving collaboration while tracking data contributions and supporting fairness mechanisms [35,36]. Other approaches rely on decentralized storage solutions (e.g., IPFS) to ensure the reproducibility and integrity of datasets, models, and code artifacts [37,38]. Overall, the field is moving beyond conceptual notions of trust toward practical and verifiable implementations, emphasizing provenance tracking, auditing, access control, and Blockchain-based governance across diverse AI application domains [38-41].

Across the 49 studies in our dataset, Blockchain is referenced in all papers (49 studies, 100%), followed by smart contracts (26 studies, 53%). DLT and consensus variants (e.g., PoW, PoS, PoA, PoT, PoFL) appear in 18 studies (36%), federated learning in 9 studies (18%), and decentralized storage or provenance mechanisms in 6 studies (12%). Advanced cryptographic techniques (e.g., HE, ZK, SMPC) and Blockchain-supported XAI each occur in 5 studies (10%). Explicit LLM integrations targeting traceability or explainability are reported in 3 studies (6%), while AI artifact tokenization (e.g., NFTs) appears in 2 studies (4%), and oracles or trusted data feeds in 1 study (2%) (Fig.18).

## RQ2.1- Type of AI Application improved

Blockchain adoption in AI research is most prevalent in multi-party settings where large volumes of data must be governed, shared, and audited [40,42]. Healthcare and IoT/AIoT emerge as the dominant application domains, reflecting strong requirements for data provenance, secure data exchange, and reliability in sensor-intensive environments [35,43,44]. Other frequently studied domains include Finance/Web 3.0 (e.g., DeFi and digital marketplaces) and Supply Chain/Logistics, where on-chain traceability directly supports accountability and governance of data and AI models [45-47]. Additional applications span Smart Cities, Energy Systems, and Robotics (Industry 4.0), targeting auditable and resilient cyber-physical operations such as safety monitoring, anomaly detection, and collaborative control [42,48]. Overall, these trends suggest that current research prioritizes accountability and verifiable data flows over usability concerns, focusing on domains where Blockchain-based provenance and access control provide tangible trust benefits [49].

Figure 19 summarizes the application domains across the 49 studies. Healthcare and IoT/AIoT/IIoT/IoV are each addressed in 16 studies (32.7%). Smart cities, energy systems, and cyber-physical systems appear in 11 studies (22.4%). Finance/Web 3.0/DeFi/DAOs/marketplaces and Robotics/Industry 4.0 are each covered by 10 studies (20.4%). Supply chain and logistics account for 7 studies (14.3%), while data sharing, data markets, and AI marketplaces appear in 5 studies (10.2%). Less frequently explored domains include cloud-centric AI operations (4 studies, 8.2%), governance and regulatory decision-making (3 studies, 6.1%), software engineering (testing/SDLC/DevOps) and education (2 studies each, 4.1%), and niche areas such as the metaverse and tourism/online reviews (1 study each, 2.0%).

**RQ2.1- Types of Trustworthiness**

The trust goals addressed by combining AI and Blockchain in the reviewed studies can be grouped into three overarching categories. The most dominant category is security and privacy, where Blockchain is used as an infrastructural control layer for AI data pipelines and models, supporting secure data sharing, access control, and resistance to cyber-attacks. The second category focuses on transparency, traceability, and auditability, typically realized through smart contracts and on-chain registries (e.g., NFTs, IPFS). These mechanisms make artifacts such as data ownership, model lineage, and decision logs observable and verifiable, thereby enabling governance, accountability, and compliance in decentralized AI settings. The third category addresses **behavioral and system-level trust**, including reliability, robustness, explainability, and fairness. In this group, Blockchain is paired with techniques such as explainable AI (XAI) and fairness-aware federated learning to support interpretable, robust, and equitable AI behavior. Consistent with earlier observations, empirical studies predominantly emphasize security-oriented trust mechanisms, whereas conceptual and survey-based works place greater emphasis on transparency, governance, and accountability.

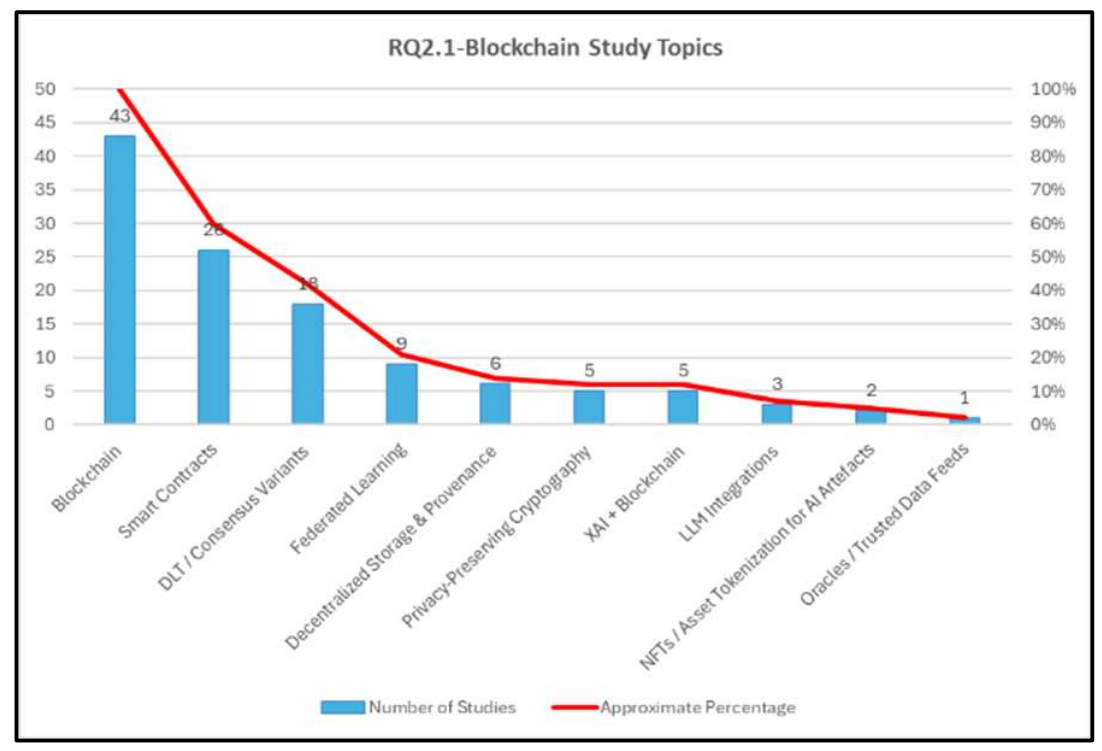

**Fig. 18** Blockchain Study Topics

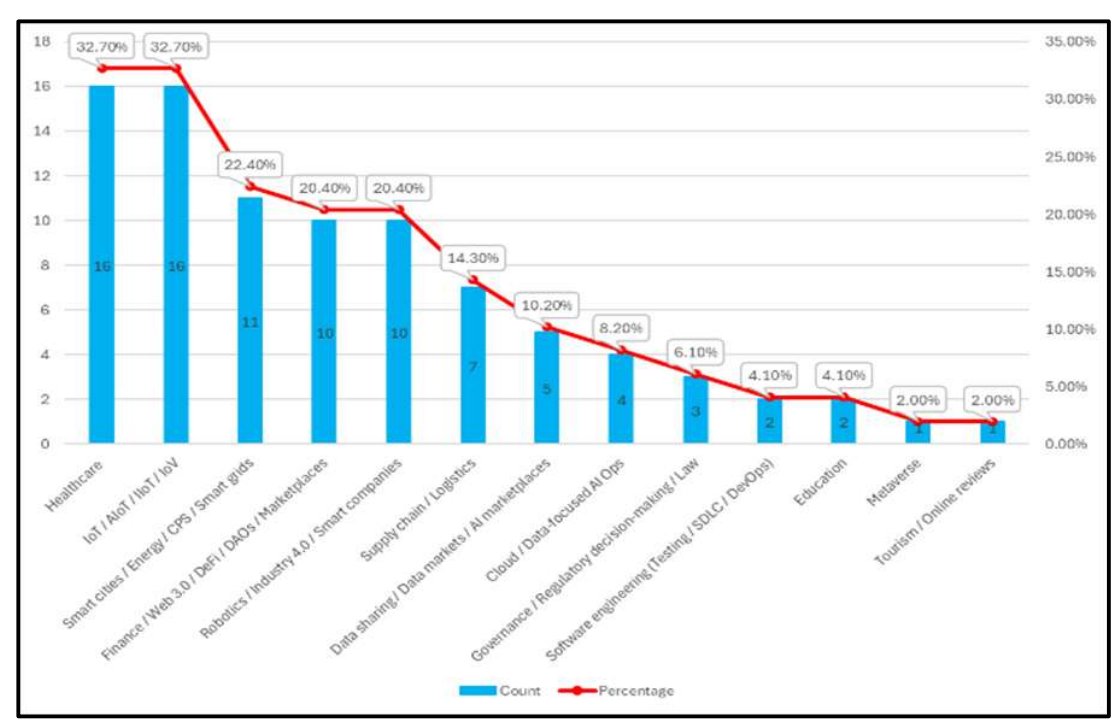

Fig. 19 Application Domains in BAISET Research

Across the 49 analyzed studies (Fig.20), security and threat mitigation are addressed most frequently (28 studies, 57.1%), followed by transparency/traceability/accountability (19 studies, 38.8%), privacy and data protection (14 studies, 28.6%), and data integrity/provenance (11 studies, 22.4%). Trust dimensions that receive comparatively limited attention include reliability and robustness (7 studies, 14.3%), explainability (XAI) (5 studies, 10.2%), fairness (4 studies, 8.2%), auditability (4 studies, 8.2%), lawfulness and compliance (3 studies, 6.1%), integrity as a standalone property (3 studies, 6.1%), and authenticity/non-repudiation (2 studies, 4.1%). This distribution mirrors the broader maturity of the field, with a strong focus on ledger-centric assurances (security, provenance, accountability) and comparatively less emphasis on user-facing and behavioral trust properties such as fairness and explainability.

**RQ2.1- Type of Evaluation**

The distribution of evaluation types indicates that blockchain-enabled trustworthiness for AI is still at an early stage of empirical maturity. Literature is largely dominated by analytical and exploratory studies, while large-scale, reproducible, and deployment-level validations remain scarce. Consistent with earlier RQs, the evidence base is skewed toward surveys and conceptual work, with relatively few quantitative or real-world evaluations.

Within the 49 studies of the RQ2.1 corpus (Fig.21), survey papers constitute the largest share, accounting for 25 studies (51%). These include scoping reviews, systematic mappings, and taxonomies across domains such as healthcare, federated learning, governance, robotics, and deep-learning systems. Empirical studies comprise 14 papers (29%), typically reporting prototype-based evaluations or controlled experiments in areas such as EHR data sharing [51], federated learning integrity [40], SDN-based DDoS mitigation [53], provenance toolchains [40], ML-based consensus mechanisms [54], and metaverse prediction services [34]. Conceptual papers account for 10 studies (20%), focusing on architectural blueprints, lifecycle models, and socio-technical trust frameworks, often without implementation or quantitative assessment.

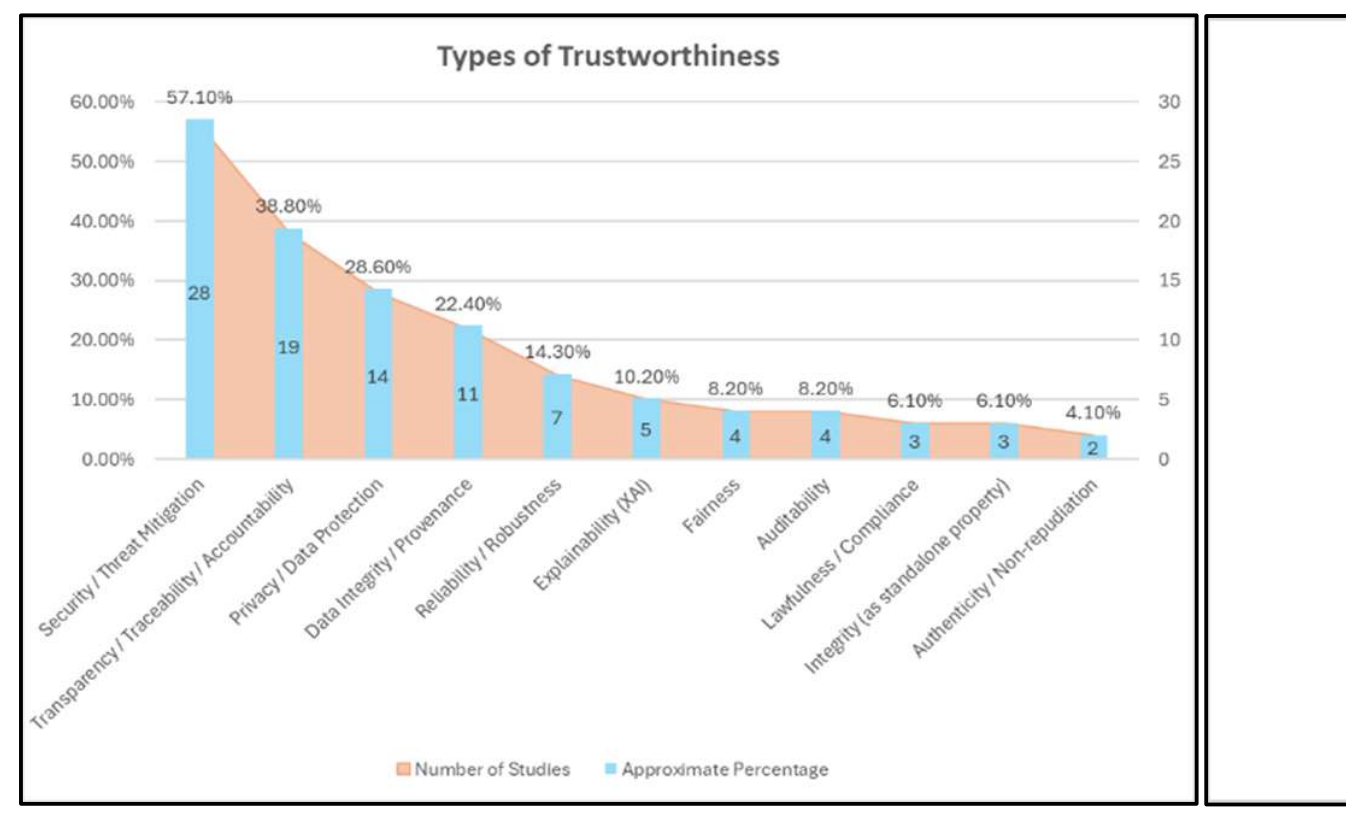

**Fig.20** Types of Trustworthiness

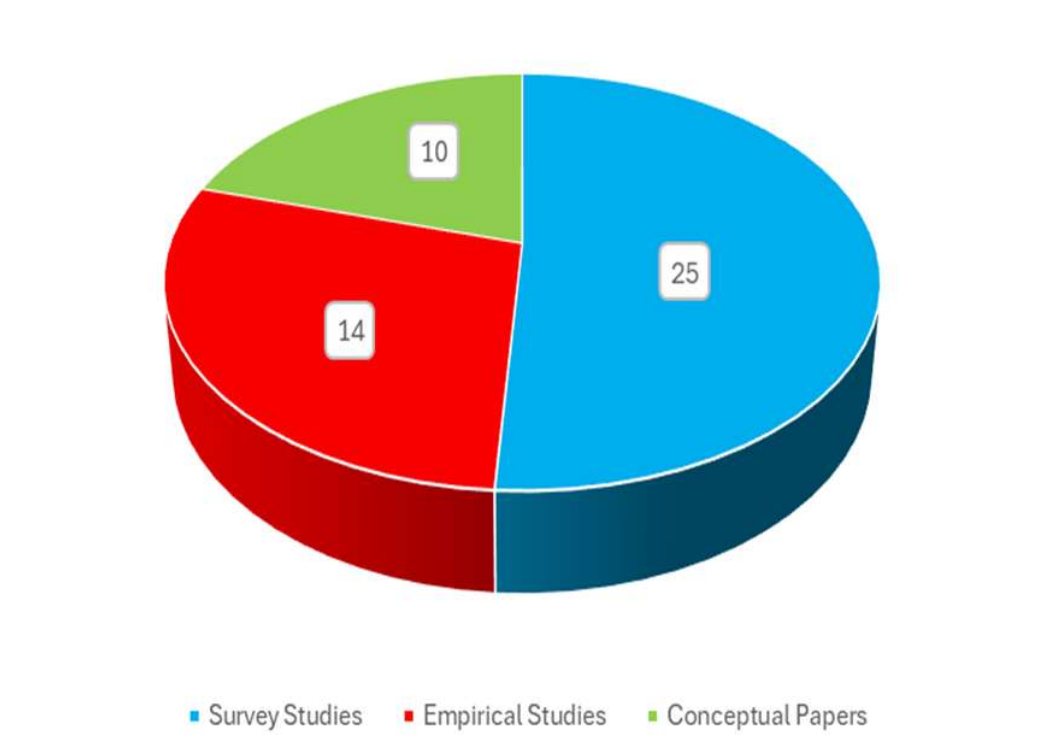

**Fig. 21** Study Type in Literature Review

Overall, the literature prioritizes problem scoping, requirement consolidation, and architectural reasoning over deployment-level validation. This mirrors broader RQ2 findings, where trust is primarily operationalized through design-time safeguards and analytical assurances rather than measured system behavior. The results expose a clear research gap advancing toward systematic empirical validation through comparative benchmarks, shared datasets, and real-world field trials to assess the scalability, effectiveness, and sustainability of Blockchain-based trust mechanisms for AI across application domains.

**Designing and planning.** Nine studies (18.4%) focus on the design and planning phase of trustworthy AI-Blockchain integration, emphasizing the conceptualization and formalization of methodological foundations. These works are evenly distributed across three subcategories: Requirements and Architecture Definition [F24, M30, M42], which propose system-level blueprints; Methodology, Process, and Workflow Design [M21, F21, F25], which formalize operational procedures and pipelines; and Governance, Policy, and Risk/Threat Modeling [M26, B7, B13], which analyze regulatory contexts and propose governance models. Together, this cluster reflects foundational design efforts aimed at structuring regulated, auditable, and methodologically sound AI-Blockchain ecosystems.

**Searching and researching.** Sixteen studies (32.6%) emphasize research-oriented investigations exploring the convergence of Blockchain and AI, reflecting the early-stage maturity of the field. These works [M1, M2, M4, M7, M9, M21, M22, M25, M27, M29, M41, M44, B1, B13, F6, F19] focus on exploratory and analytical contributions rather than concrete implementations. Their primary objective is to examine conceptual frameworks, analytical taxonomies, and empirical observations related to trust mechanisms, transparency, and explainability in decentralized AI ecosystems. This cluster highlights Blockchain's role in securing AI data, models, and decision processes at a conceptual level.

**Content generation.** A majority of the studies 31 papers (63.3%) report producing tangible outputs that operationalize trustworthy AI-Blockchain integration. The most common outputs are Tools, Platforms, Prototypes, or Systems (21 studies) [M22, M24, M25, M27, M30, M31, M42, B4, B7, F1, F8, F9, F12, F14, F18, F21, F24, F27, F28, F29, F30], delivering functional implementations that combine AI analytics or federated learning with blockchain-based audit and provenance mechanisms. This is followed by Framework, Model, or Method Proposals (7 studies) [M24, M25, M30, F9, F21, F24, F25], which formalize governance and lifecycle blueprints. Less frequent outputs include Protocols or Smart-Contract Templates (2 studies) [B4, F1] and Guidelines or Taxonomies (1 study) [B13]. Overall, this cluster underscores a growing emphasis on executable artifacts that directly support transparency, auditability, and trust.

**Analysis and application.** Fourteen studies (28.6%) focus on analysis and application-oriented evaluation, including diagnostic validation, vulnerability analysis, and reliability assessment of AI-Blockchain systems. These works primarily conduct post-deployment evaluations to identify faults, assess resilience, and validate smart-contract behavior. The analyses target security and robustness testing (10 studies) [M4, M5, M9, M22, M28, M29, M42, B2, F24, F27] as well as auditability and trust-consistency checks (4 studies) [M31, M43, B7, F3]. This cluster highlights the essential role of systematic analysis in achieving resilient, transparent, and fault-tolerant AI-Blockchain ecosystems.

| **RQ2.1** How has Blockchain being applied to improve trustworthiness in AI-related applications? |
| --- |
| Our review of 49 studies shows that AI-Blockchain integration for trustworthiness is most commonly realized through smart contracts, which act as the primary enforcement mechanism. The technology stack is heterogeneous, frequently combining DLT/consensus mechanisms, federated learning, and decentralized storage (e.g., IPFS). Healthcare and IoT/AIoT dominate application domains (each 33%), reflecting strong demand for secure, multi-party data governance. Trust objectives focus mainly on Security (57%) and Privacy (29%), with moderate attention to Transparency/Traceability (39%) and Data Integrity/Provenance (22%), while Fairness and Explainability remain underexplored. Methodologically, the field is largely exploratory, dominated by survey-based studies (51%), with limited empirical prototype evaluations (29%) and scarce real-world validation. |

## RQ2.2 Has Blockchain been widely adopted to improve trustworthiness in SE related applications?

### RQ2.2-Technology

Among the 26 studies focusing on SE, Blockchain serves as the foundational technology for establishing trust controls, with smart contracts providing the primary mechanism to enforce these controls, for example, in requirements tracking, continuous delivery governance, and auditing. As noted, "knowledge on technical and non-technical means to realize TAI is scattered across disciplines, making it challenging to grasp the status quo on its realization" [55]. Many studies combine Blockchain with AI/Machine Learning techniques, such as LSTMs or XAI, primarily to support security testing, model and data lineage management, and governance. LLMs are emerging, mostly for code generation, automation, or assisting human governance processes. Decentralized storage solutions, such as IPFS, are also common, while specific ledger types (DLTs), vendor platforms (e.g., Hyperledger), and advanced cryptography (e.g., PoFL, TEEs) are applied to meet specialized requirements, including compliance, privacy-preserving model training, and device trust. Collectively, these studies adopt a "platform-plus-glue" approach, with Blockchain as the core backbone for auditing and provenance, complemented by AI, decentralized storage, and other tools to address SE tasks across the software lifecycle [56, 57].

Across the 26 SE-focused studies, Blockchain dominates (25 studies, 96.2%), with Smart contracts widely used (19 studies, 73.1%) to operationalize trust mechanisms. AI/ML techniques, including XAI, agents, and federated learning appear in 11 studies (42.3%), often supporting security and governance tasks. Decentralized storage (IPFS) appears in 6 studies (23.1%), while DLT/ledger variants, tokens/NFTs/DAOs, and TEEs or other cryptographic privacy techniques each appear in 4 studies (15.4%). Less frequently applied are consensus mechanisms (11.5%), Hyperledger (11.5%), Ethereum (7.7%), federated learning (7.7%), and LLMs (7.7%). Single studies (3.8%) cover provenance standards or SE-specific methods/tools (e.g., UML, Agile). Overall, the landscape reflects a "platform-plus-glue" model, where Blockchain provides the central trust backbone, complemented by AI, storage, and specialized tools to address diverse SE tasks (Fig.22).

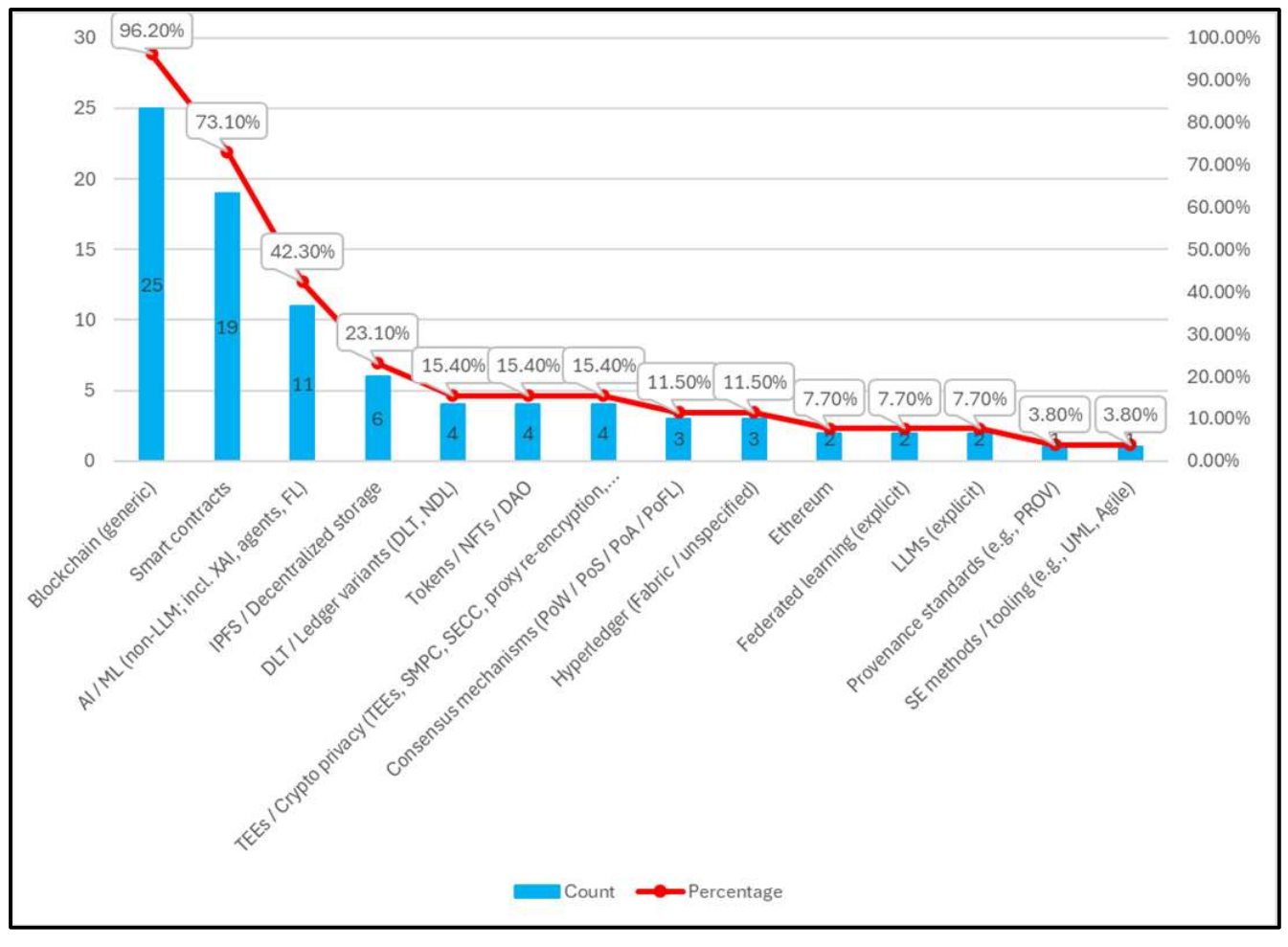

**Fig. 22** Technologies Studied

### RQ2.2 - Type of SE Task Being Supported

In our dataset, Blockchain is primarily employed as a robust infrastructure for governing data, managing models, and enforcing compliance, rather than as a simple coding tool [58]. Over half of the research focuses on two main areas: data and model management (provenance, versioning, access) and governance, compliance, and auditing [41]. This reflects a pragmatic approach aimed at making AI and software assets traceable, verifiable, and accountable across their lifecycle [32]. Security and privacy are generally embedded within operational workflows such as device trust or secure review rather than treated as standalone functions [40]. Applications in Testing/QA and Development/CI are present but limited, while support for Requirements/Design is rare, underscoring a need for earlier-stage tooling. Overall, Blockchain operationalizes trust around processes, data, and models, not merely network endpoints [56].

Across the 26 studies, the distribution of SE tasks supported by Blockchain is as follows: Data and model management appears in 14 studies (53.8%), Governance/compliance/audit/risk/accountability in 10 (38.5%), Security and privacy in 9 (34.6%), Deployment and operations/DevOps in 7 (26.9%), Explainability/transparency/fairness in 5 (19.2%), Testing/verification/QA in 4 (15.4%), Development/CI/program repair in 4 (15.4%), Requirements/design/traceability/change management in 3 (11.5%), and Collaboration/decision-making/consensus in 3 (11.5%)

### RQ2.2 - Type of Trustworthiness

Among the 26 SE-focused studies, transparency is the most frequently addressed trustworthiness dimension, reported in 25 studies (96%) and typically implemented via ledgers for audit trails, decision traceability, and accountability. Integrity appears in 20 studies (77%), followed by robustness in 15 (58%). Accountability and reliability each occur in 13 studies (50%), while security and privacy are reported in 12 studies (46%). Less frequently addressed dimensions include lawfulness/ethicality/compliance (5 studies, 19%), fairness/justice and immutability (4 studies each, 15%), explainability/explicability (3 studies, 12%), and anonymity (2 studies, 8%). Traceability and other singletons are reported in only 1 study each (4%). The limited coverage of user-centered attributes fairness, explainability, immutability, anonymity, and other singletons highlights a gap beyond ledger-focused assurances such as provenance and immutability (Fig.23).

#### RQ2.2 -Type of Evaluation

From the 26 studies analyzed, surveys and empirical evaluations are equally prevalent (10 studies each, 38.5%), with conceptual work accounting for the remainder (6 studies, 23.1%). This balance suggests a maturing field in which evidence collection and prototype-based validation are increasingly common. However, the continued dominance of surveys over controlled experiments indicates a reliance on perception-driven and context-specific evaluations, often based on non-standardized datasets, which constrains benchmarking and cross-study comparability (Fig. 24).

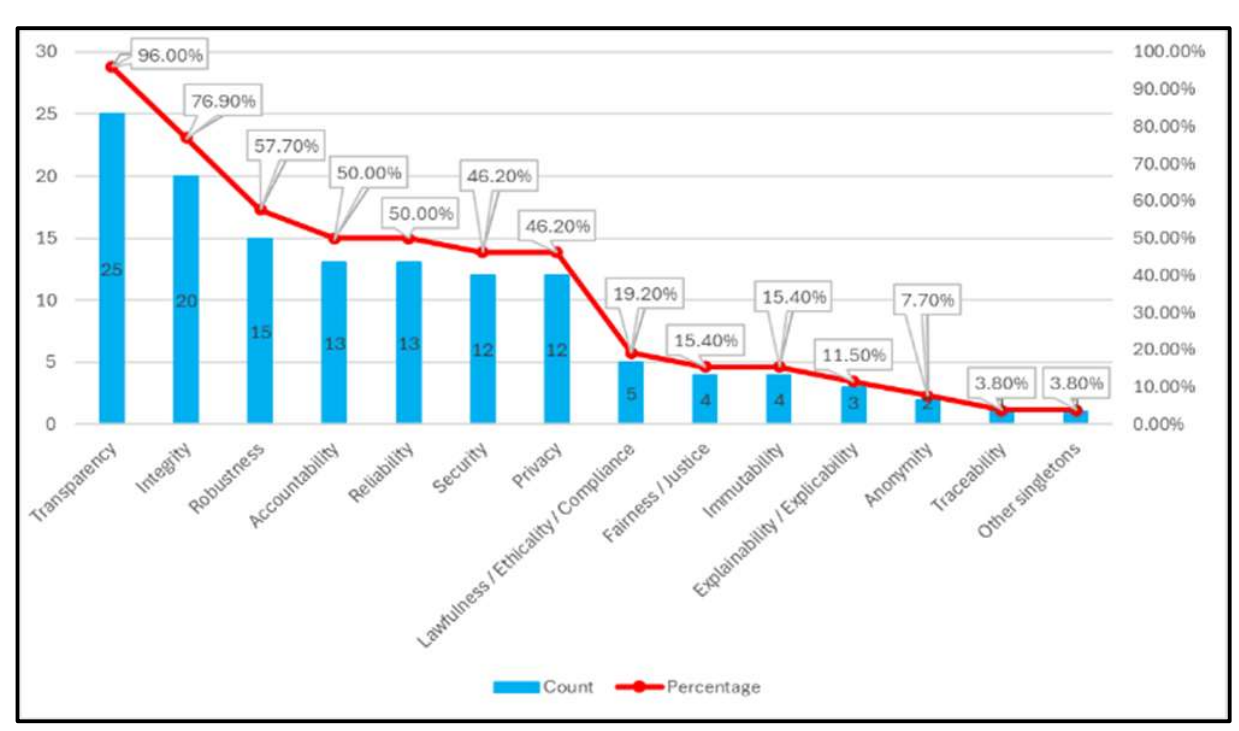

**Fig. 23** Distribution of Trustworthiness Attributes Blockchain-SE Research

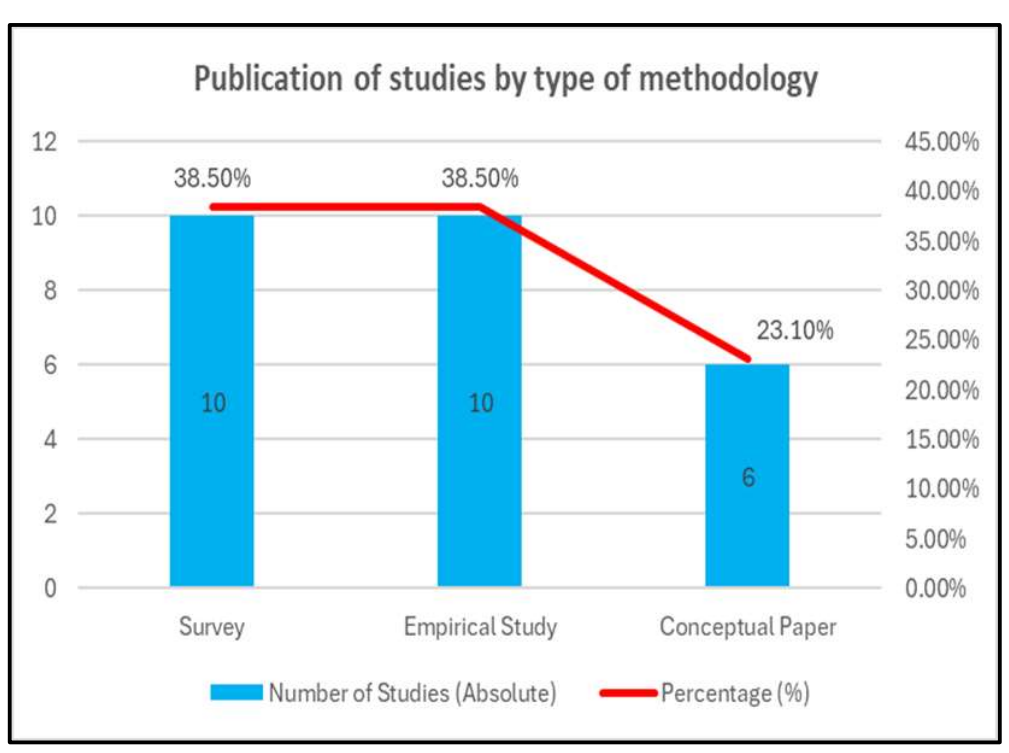

**Fig. 24** Publication of studies by type of methodology

**Designing and Planning:** Of the 26 papers reviewed, 8 studies (30.8%) focus on evaluation design and planning, detailing how architectures, frameworks, and governance mechanisms are conceptualized to enhance trust and transparency. These are grouped into three categories: Requirements and Architecture Definition (3 studies [M14, F14, F30]) establishing structural blueprints and system requirements; Methodology/Process/Workflow Design (3 studies [M19, M21, M39]) formalizing operational procedures and data lifecycles using federated learning and blockchain provenance; and Governance/Policy/Risk & Compliance Modeling (2 studies [M42, B7]) emphasizing accountability, ethical oversight, and audit mechanisms. This cluster highlights foundational work essential for operationalizing secure, explainable, and regulation-aligned AI-Blockchain ecosystems.

**Searching and Researching:** Nine studies (34.6%) address exploratory and analytical investigations into Blockchain-AI convergence ([M19, M21, M39, M41, M43, B1, B2, F19, F25]). These works primarily examine conceptual frameworks, governance models, and analytical taxonomies defining trust relationships between the two technologies. They emphasize transparency, explainability, accountability, and privacy, showing how Blockchain ensures data provenance and auditability while AI enhances security and decision intelligence. This cluster reflects the analytical maturity of the field by mapping theoretical and early-stage models of trustworthy AI-Blockchain systems.

**Content Generation:** Fifteen studies (57.7%) report tangible outputs operationalizing AI-Blockchain integration. The largest category is Tools, Platforms, Prototypes, or Systems (9 studies: [M13, M22, B8, F14, F18, F24, F27, F29, F30]), delivering functional implementations for decentralized auditing, secure pipelines, and federated data integrity. Frameworks, Models, **or** Method Proposals (5 studies [M14, M39, B1, F21, F30]) provide structured blueprints for governance, ethical compliance, and federated learning. Guidelines and Taxonomies appear in 1 study ([F25]), offering conceptual and ethical design references. This cluster underscores the field's focus on executable artifacts and formal models that enhance transparency, accountability, and auditability.

**Analysis and Troubleshooting:** Ten studies (38.5%) focus on post-deployment analysis, assessing faults, verifying security controls, and evaluating reliability. Two main dimensions are observed: Security and Robustness Testing (5 studies [M22, M41, F14, F24, F28]), addressing system resilience, vulnerability mitigation, and smart-contract validation; and Auditability and Trust-Consistency Checks (5 studies [M39, M43, B7, B8, F30]), focusing on decentralized auditing, compliance monitoring, and accountability tracing. These works highlight the critical role of systematic analysis in building resilient, transparent, and verifiable AI-Blockchain ecosystems.

#### RQ2.3 Has Blockchain being used to improve the trustworthiness of AI-based SE approaches

#### RQ2.3- Technology used

The RQ2.3 dataset comprises 17 papers on AI-based software evolution (SEvo). Blockchain is consistently employed, typically via smart contracts. Two dominant approaches emerge: (1) Provenance and Asset Governance, where smart contracts are combined with tools like IPFS to version and track models, datasets, and code; and (2) Secure or Fair Learning, where smart contracts integrate with Federated Learning (FL) to enforce accountability and reinforce training processes. Other blockchain platforms (e.g., DLTs) appear in a smaller subset, while specialized trust tools such as NFTs and TEEs are mostly applied in marketplace and access-control contexts. On the AI side, most studies pair blockchain with ML/DL, with a few exploring agents or LLMs for decentralized coordination. Overall, the technology stack reflects a shift from conceptual reviews toward practical, implementable lifecycle control for data and models.

Across the 17 studies, the frequency of technologies is as follows: Blockchain in 16 studies (94.1%) and Smart Contracts (including chaincode) in 12 studies (70.6%) dominate; SSI and ML/DL components appear in 9 studies each (52.9%). Federated Learning occurs in 6 studies (35.3%) and IPFS in 5 studies (29.4%), followed by permissioned ledgers in 3 studies (17.6%). Agents/LLMs and GAN/CNN/RNN each appear in 2 studies (11.8%), while NFTs, DAO/DeFi, TEEs, proxy re-encryption, provenance graphs, and IBM BaaS each appear in 1 study (5.9%) (Fig. 25).

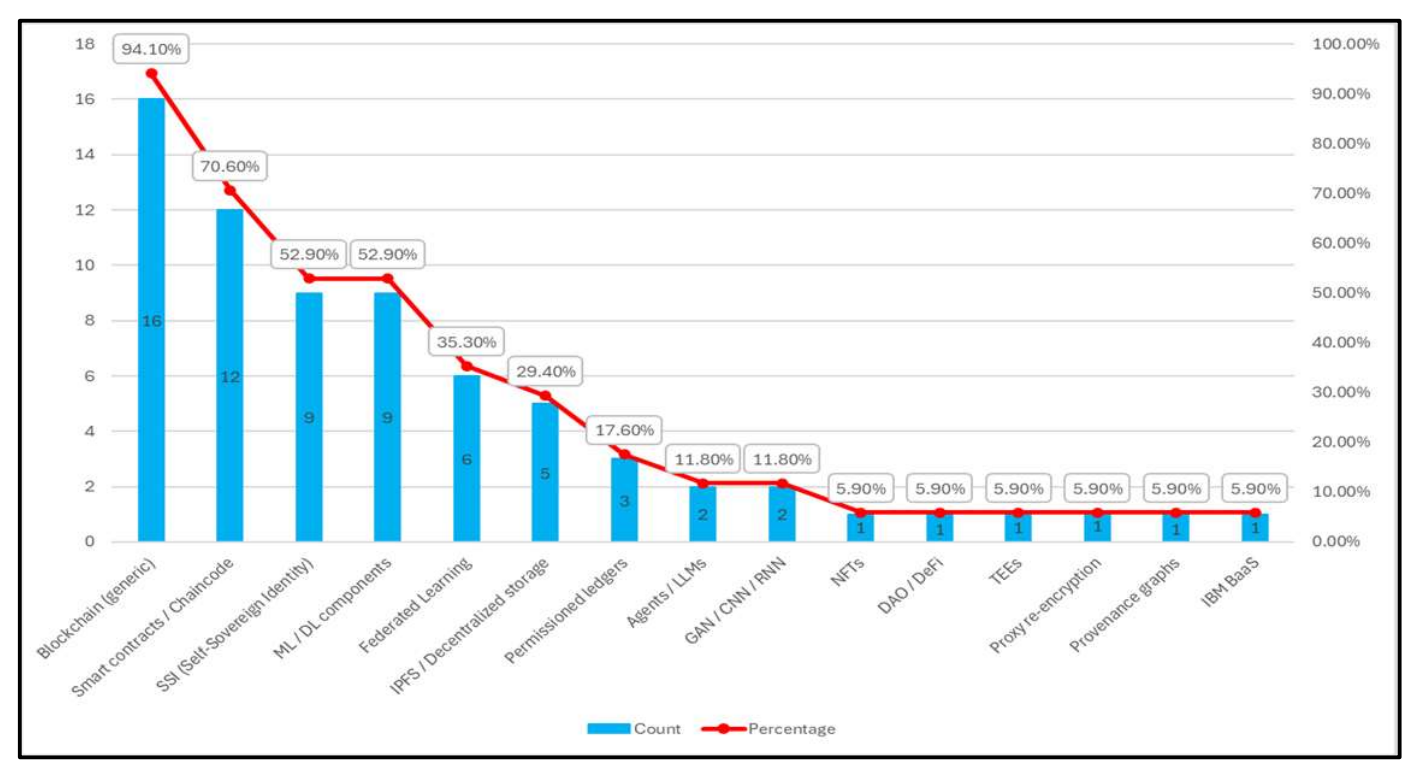

**Fig.25** Distribution of technologies used

**RQ2.3- Type of Software Evolution task being supported Type of SE task being supported**
In the context of software evolution, Blockchain is primarily used to enforce governance of data and models (e.g., tracking provenance, versioning, and access control) and to mitigate operational risks such as security monitoring and vulnerability detection. Training support is generally included only for federated or secure learning scenarios. Deployment and operations tasks (e.g., access control, oversight) and testing, verification, and collaboration (e.g., shared updates) appear less frequently. Uses such as marketplaces, asset trading, or pure decision-support are rare. Overall, the distribution of tasks indicates a shift from early security-focused applications toward full-lifecycle control, aiming to make AI artifacts and software processes traceable, auditable, and accountable throughout development and deployment.

Across the 17 studies, data and model management is the most common focus (13 studies, 76.5%), followed by governance and compliance (10 studies, 58.8%) and security monitoring (9 studies, 52.9%). Training and access control each appear in 4 studies (23.5%), while testing, reproducibility, and collaboration appear in 3 studies each (17.6%). Marketplace functions are reported in 2 studies (11.8%), and decision-making tasks are least frequent (1 study, 5.9%) (Fig. 26).

**RQ2.3- Type of trustworthiness**
For software evolution, trust shifts from merely hardening endpoints (e.g., firewalls) to ensuring that all artifacts and processes are observable and tamper-proof. Transparency is the dominant focus, almost always paired with integrity and accountability, reflecting heavy reliance on ledgers and smart contracts to enforce audit trails, data provenance, and rules. Privacy is also common, often implemented alongside technologies such as federated learning (FL) and IPFS. Fairness appears in about half the studies, typically addressed via data weighting or governance mechanisms during training. Security and reliability are present but less dominant, usually embedded within operational workflows rather than treated as standalone objectives. Only a few studies consider ethics/compliance and explainability, highlighting an opportunity to align technical assurances with human-centered trust requirements.

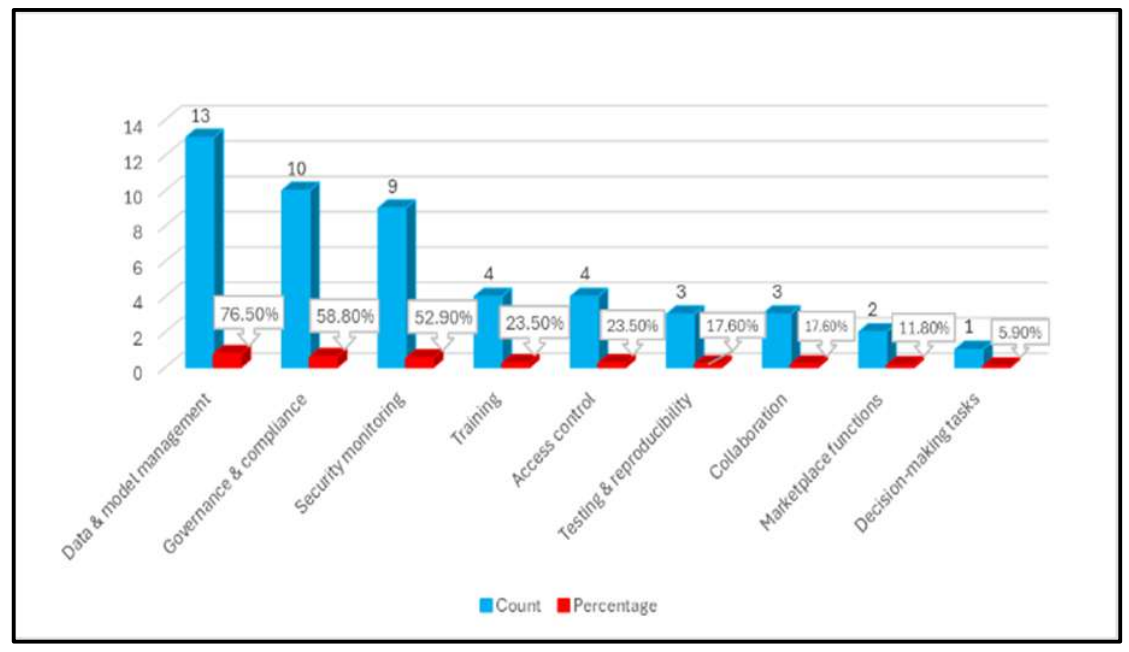

**Fig.26** Distribution of Type of Software Evolution task

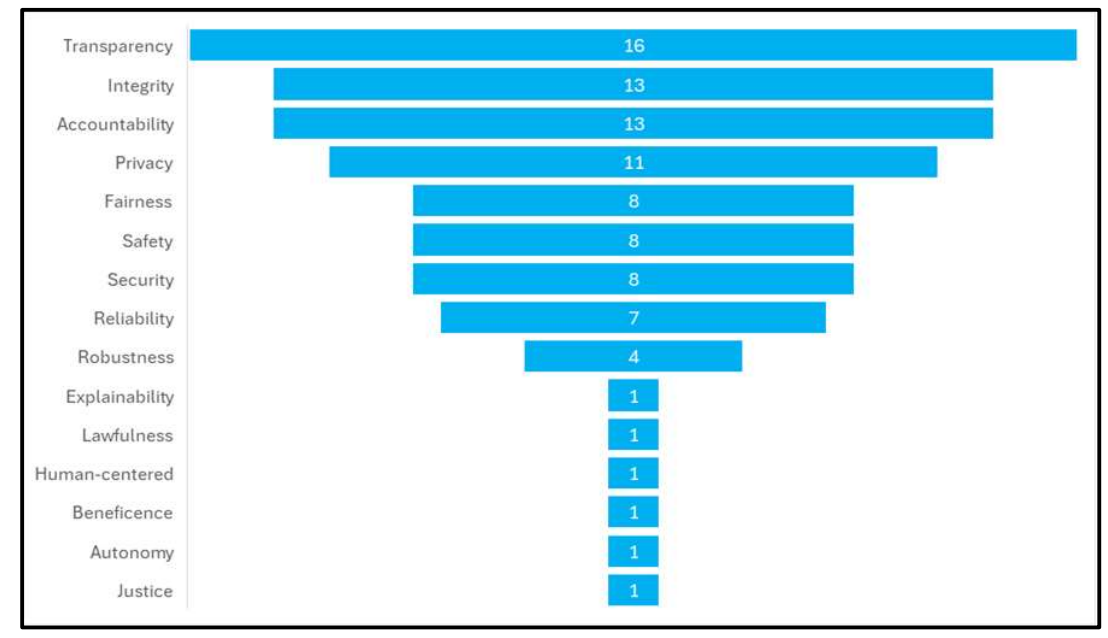

**Fig.27** Trustworthiness type distribution

Among the 17 studies, transparency appears in 16 studies (94.1%), integrity and accountability each in 13 studies (76.5%), privacy in 11 studies (64.7%), and fairness, safety, and security each in 8 studies (47.1%). Less frequently addressed dimensions include reliability in 7 studies (41.2%) and robustness in 4 studies (23.6%). Other criteria explainability, lawfulness, human-centeredness, beneficence, autonomy, and justice appear only once each (5.9%) (Fig. 27).

**RQ2.3- Type of Evaluation**
**Evaluation Types:** Surveys are the most common evaluation type, followed by empirical prototypes or case studies, and conceptual proposals. This indicates that the field is progressing from theoretical analysis toward empirical validation. However, large-scale, comparative, and long-term evaluations remain scarce. There is a continuing need for benchmarkable tasks, shared datasets, and replicable pipelines that consistently report both effectiveness and associated overhead (e.g., latency or energy consumption). Among the 17 mutually exclusive studies, surveys appear in 8 studies (47.1%), empirical studies in 5 studies (29.4%), and conceptual papers in 4 studies (23.5%) (Fig. 28).

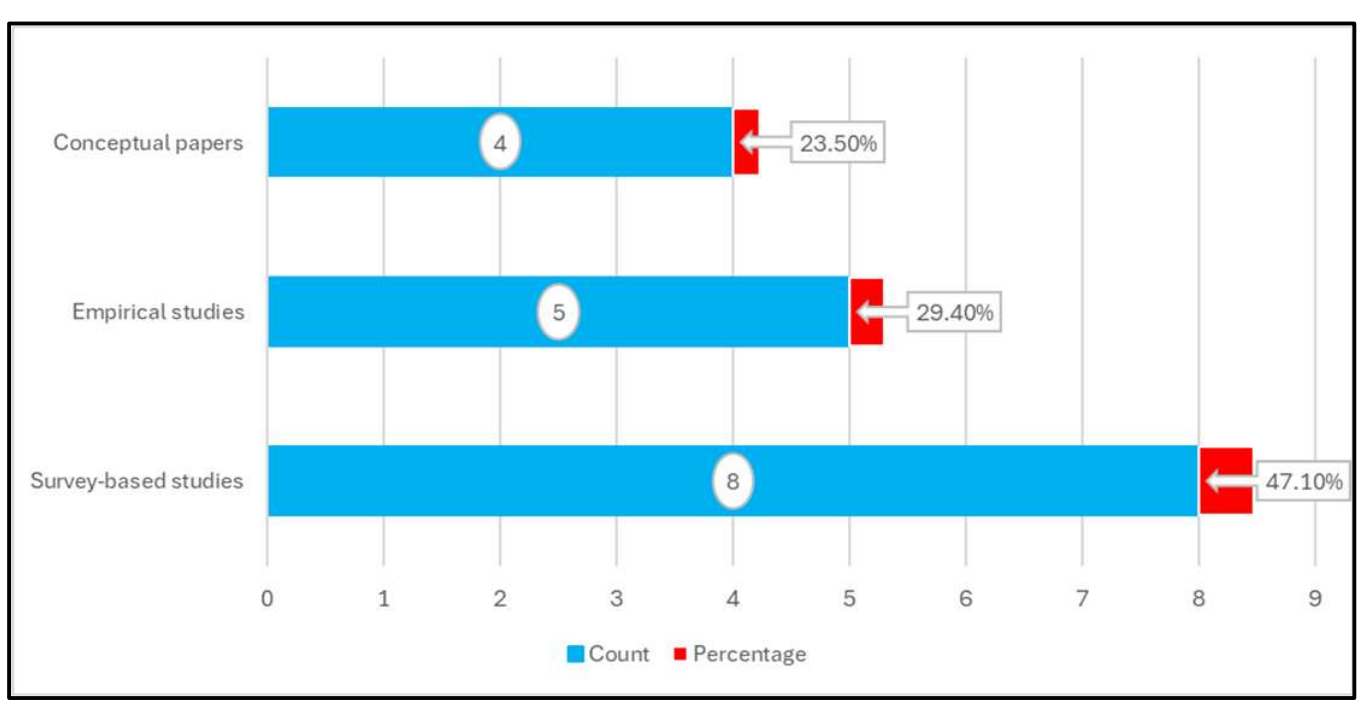

**Fig.28** Evaluation type distributions

**Designing and Planning:** Seven studies (41.2%) focus on the design and planning phase, detailing architectures, workflows, and governance mechanisms for trustworthy AI-Blockchain integration. These are grouped into three categories: Requirements and Architecture Definition (3 studies [M21, M27, M39]) establishing structural blueprints; Methodology/Process/Workflow Design (3 studies [M21, M31, F18]) formalizing operational workflows; and Governance/Policy/Risk & Compliance Modeling (1 study [M43]) addressing ethical oversight and regulatory requirements. This cluster underscores foundational work critical for operationalizing secure, explainable, and regulation-aligned AI-Blockchain ecosystems.
**Searching and Researching:** Seven studies (41.1%) explore research-oriented investigations into Blockchain's role in trustworthy AI-based software evolution [M21, M27, M42, M43, B1, F19, F21]. These works focus on conceptual and analytical contributions rather than implementations, examining theoretical frameworks, governance models, and trust relationships. Key themes include transparency, accountability, privacy, and fairness, demonstrating how Blockchain ensures provenance and auditability while AI supports automation and adaptive learning. Collectively, these studies reflect the field's early analytical maturity.
**Content Generation:** Eleven studies (64.7%) report tangible outputs that operationalize trustworthy AI-Blockchain integration. The largest group is Tools, Platforms, Prototypes, or Systems (6 studies [M30, M31, F18, F29, F30, M42]), delivering functional implementations for secure data sharing, provenance tracking, and federated learning. Frameworks, Models, or Method Proposals (4 studies [M27, M39, M42, F30]) outline structured blueprints for decentralized governance, fairness, and ethical compliance. Guidelines and Taxonomies appear in 1 study ([F25]), providing conceptual and regulatory guidance. This cluster highlights the emphasis on creating executable artifacts that strengthen transparency, accountability, and trust.
**Analysis and Troubleshooting:** Eight studies (47%) focus on post-deployment analysis and troubleshooting [M28, M29, M30, M31, M42, M43, F18, F28], assessing reliability, verifying security mechanisms, and evaluating resilience in AI-Blockchain systems. Two main evaluation dimensions are identified: Security and Robustness Testing (5 studies [M30, M31, M43, B1, F28]), addressing vulnerability detection, smart-contract validation, and model reliability; and Auditability and Trust-Consistency Checks (3 studies [M39, M43, F18]), emphasizing decentralized auditing, accountability monitoring, and provenance verification. These evaluations are critical for ensuring resilient, transparent, and verifiable Blockchain-AI ecosystems.

| **RQ2.3** Has Blockchain being widely adapted to improve trustworthiness in SE related approaches? |
| --- |
| Smart contracts dominate Blockchain-enabled AI applications (70%), with research primarily targeting Data & Model Management (76.5%) and Governance/Compliance (58.8%). Transparency is nearly universal (94.1%), while Integrity and Accountability are also strongly considered (76.5%). Fairness and Privacy receive moderate attention, but human-centric aspects like Explainability and Ethics remain underexplored. Evaluations are largely survey-based (47%), with empirical studies less common (29%), reflecting the field's early-stage validation. |

RQ3- **Type of LLMs**

Across the 4 studies included in the RQ3 dataset, most papers address the governance and provenance of LLMs for software engineering (LLM4SE) at a tool-agnostic level, without benchmarking specific models. When particular models are referenced, the focus is on widely used coding and content-generation services such as ChatGPT and GitHub Copilot, with occasional mentions of Codex, Gemini, and MidJourney. In general, these studies discuss the ways in which blockchain can support traceability, integrity, and policy enforcement in LLM-driven SE workflows. Specific references within the 4 non-exclusive studies include general LLMs [F12, F13], ChatGPT [M44, F13], GitHub Copilot [F13, M44], and single mentions of Codex [M44], Gemini [F13], and MidJourney [F13]. Overall, the focus remains on conceptual frameworks and governance mechanisms rather than empirical evaluation of LLM performance.

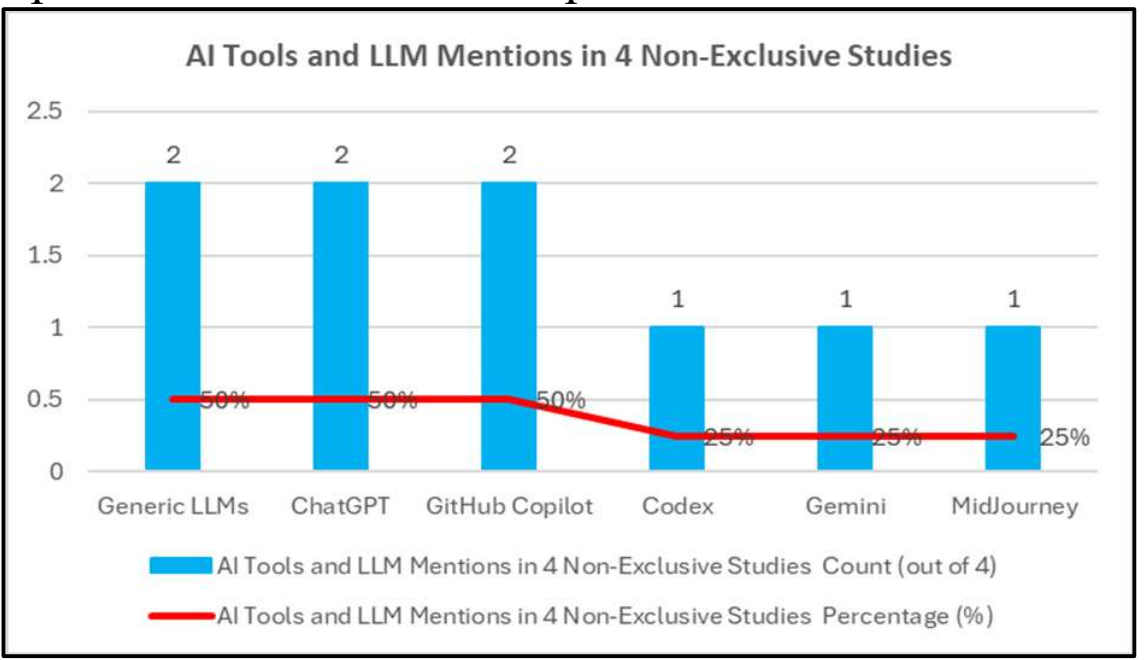

**Fig. 29** AI Tools and LLM Mentions in 4 Non-Exclusive Studies

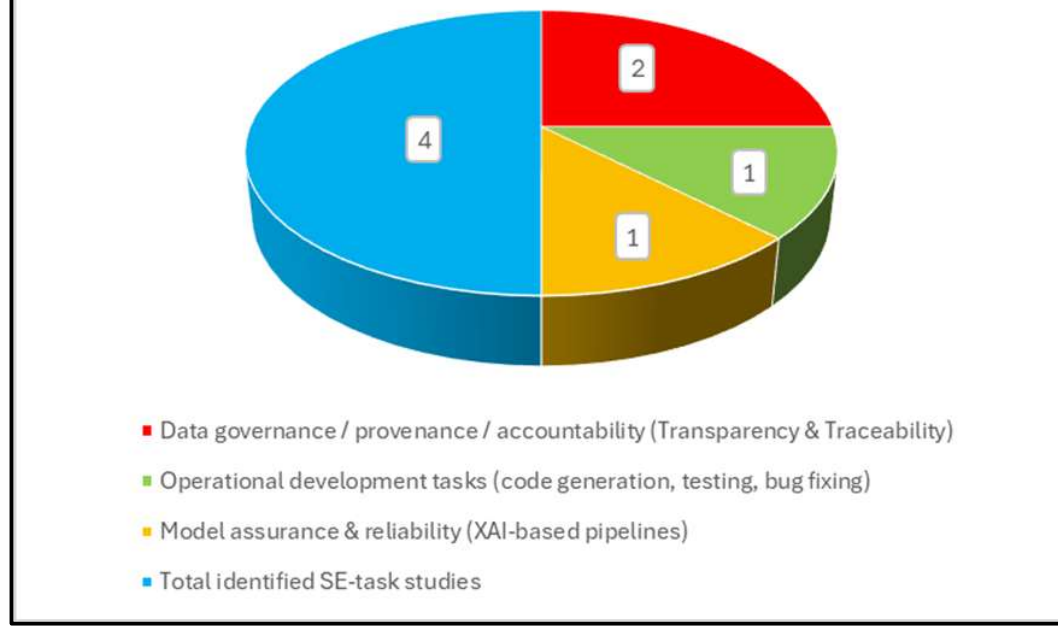

**Fig. 30** Supported SE Task Type Distribution

RQ3- **Type of SE task being supported**

Among research combining LLMs with blockchain for SE tasks, most studies focus on governance over LLM artifacts and processes. Blockchain is used to make LLM content and decisions verifiable, accountable, and policy-compliant, rather than solely supporting automated coding. Only one paper [M44] targets the direct coding workflow (code and test generation), while two papers [M44, F12] emphasize provenance, accountability, and data governance for LLM outputs and pipelines. Another study focuses on model assurance, including explainability and reliability specifically in healthcare. Overall, blockchain serves as the trust backbone, recording provenance and enforcing accountability for LLMs.

Among the four reviewed AI-Blockchain studies for SE [M44, F12, F13, F32], two studies [F12, F13] address data governance, provenance, security, and accountability, emphasizing transparency and traceability of AI-generated content. One study [M44] addresses operational development tasks, such as code generation, test-case design, and bug fixing, while one study [F32] focuses on model assurance, reliability, and healthcare diagnosis through XAI-based pipelines (Fig. 30).

RQ3- **Training Data**

Across the four RQ3 studies, training data is rarely and inconsistently specified. Only one study [M44] explicitly reports creating a study-constructed dataset derived from collected publications: "We create a dataset with the following data: authors, year of publication, RQs posed, methodology…". The remaining studies [F12, F13, F32] focus on governance and provenance or highlight data risks, such as bias and outdated information, without detailing the underlying corpus.

In practice, while most proposals describe mechanisms to make data lineage auditable, crucial details on data sources, scope, and quality controls remain underreported. This indicates an early stage of dataset standardization and transparency practices for AI-Blockchain integration. Moving forward, the field would benefit from standardized datasets covering diverse domains, sources, and licenses, along with on-chain registry mechanisms to ensure end-to-end verifiability of training provenance (Fig. 31).

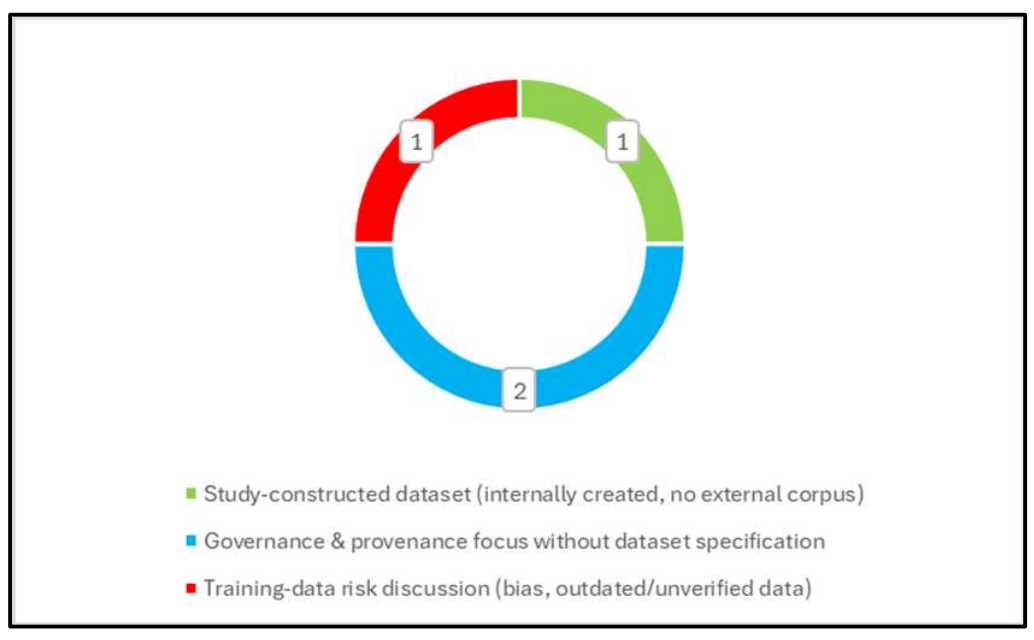

**Fig.31** Distribution of Datasets Focus on Survey

**RQ3- Type of Evaluation**

All papers in the RQ3 set are evaluated using related work surveys; none of them report on controlled experiments or deployed case studies, indicating that current evaluations of blockchain for LLM4SE are mainly conceptually (covering governance, provenance models, risks, and requirements) and no empirical evaluations are performed. This highlights the need for prototype validation using replicable tasks and performance metrics.
Across all four survey-based studies, we observe a variety of topics, including healthcare/XAI+Blockchain [F32], governance of LLMs [F12], expert insights on Copilot and ChatGPT [M44], and provenance/accountability in GenAI content [F13]. However, these papers lack empirical evaluations or provide only conceptual evaluations.

**Designing and planning.**Among the four papers reviewed, 3 papers [M44, F12, F13], focused on the design and planning phase for integrating Blockchain with LLM4SE. The goal of this research is to enhance trust and reliability through architectural frameworks and governance models. These papers propose complementary evaluation strategies [M44] focuses on architectural safeguards against data bias and outdated information in LLM training (like ChatGPT and GitHub Copilot); [F12] introduces the BC4LLM framework to formalize provenance and data integrity via blockchain-led governance; and [F13] explores design principles for responsible AI content generation, emphasizing transparency, traceability, and accountability. Overall, these studies demonstrate fundamental design efforts that are crucial for defining reliable, auditable, and ethically aligned integration between Blockchain and LLM4SE

**Searching and researching.** Of the four papers reviewed, 3 articles [M44, F12, F32] highlight how research related to Blockchain can contribute towards trustworthy LLM4SEV. These papers focus on conceptual and analytical contributions rather than practical implementations, by emphasizing theoretical exploration and conceptual modeling to reinforce transparency, accountability, and data provenance. Specifically, [M44] investigates risks and the need for verifiable provenance in AI-assisted development tools (like ChatGPT and GitHub Copilot); [F12] examines data governance through the BC4LLM framework to guarantee integrity and auditability of model updates; and [F32] explores Explainable AI (XAI) integration with Blockchain in healthcare to enhance transparency and reliability. Overall, this cluster demonstrates an early analytical maturity of Blockchain-LLM research related to LLMSEV.

**Content generation.** One study [F32] reports producing tangible outputs that illustrate how blockchain can support LLM-based software evolution (LLM4SEV) tasks, while the remaining studies [M44, F12] provide survey-based and conceptual perspectives rather than executable implementations. Paper covering tools, platforms, prototypes, or systems [F32] introduce applied mechanisms for linking blockchain with AI-assisted model validation and explainable healthcare diagnostics to enhance transparency and reliability. The remaining papers are divided between the BC4LLM Framework or Models [F12], which establishes verifiable data provenance and governance for trusted LLM pipelines, and Conceptual Guidelines [F13], these papers show how creating executable models and verifiable mechanisms can reinforce transparency, accountability, and trust in Blockchain-LLM ecosystems.

**Analysis and troubleshooting.** Of the four papers reviewed, 2 studies [M44, F32], focus on Analysis and Troubleshooting and discuss evaluation, validation, and risk mitigation in Blockchain-LLM integrations. These works, investigate how Blockchain can improve the robustness and reliability of LLM-based systems through enhanced transparency, immutability, and verifiable provenance. They cover Security and Robustness Testing [F32], which evaluates model reliability, threat detection, and data integrity within Explainable AI (XAI) healthcare frameworks; and Auditability and Trust-Consistency Checks [M44], which identify vulnerabilities and data bias in LLM-assisted coding environments like ChatGPT and GitHub Copilot, recommending continuous provenance monitoring. Collectively, these studies underscore the critical role of systematic troubleshooting in ensuring resilient, transparent, and verifiable Blockchain-LLM ecosystems.

| **RQ3.** How is Blockchain used to support LLM4SE in general and more specifically software evolution? |
| --- |
| Current research on blockchain for LLM4SE remains largely conceptual and survey-based, with no studies reporting empirical experiments or deployed prototypes. The primary focus is on trust and accountability for LLM artifacts and processes, emphasizing governance, provenance, and policy compliance. While services such as ChatGPT and GitHub Copilot are occasionally discussed, most studies concentrate on generic frameworks to ensure traceability and integrity of LLM outputs and pipelines. A notable gap is the lack of transparency regarding training data only one study explicitly reports a dataset, underscoring the need for standardized data disclosure and on-chain registries to enable full end-to-end verification. |

### RQ4. What are the limitations and recommendations for future research for the use of BAISET?

Among the 53 papers in the RQ4 corpus, the primary limitations of current Blockchain-AI Software Engineering Trust (BAISET) approaches relate to throughput, latency, and cost, often stemming from poor scalability, inefficient consensus mechanisms, and storage or gas overhead [M9, M19, M25, M28, F9, F14, F24, F27]. As noted, "increasing the capacity of Blockchain systems often comes at the cost of decreased security or decentralization, a trade-off known as the scalability trilemma" [M4].

Other major limitations include privacy and security risks, such as data poisoning and DDoS attacks [M4, M27, M30, B8, F6, F24, F30], governance gaps (interoperability, standards, regulatory alignment) [M2, M26, M28, M43, B13, F19, F21], and energy inefficiency, particularly in PoW-based systems or when integrating resource-intensive AI models [M5, M9, M13, M25, M29, F8, F12]. "Blockchain, especially proof-of-work-based systems like Bitcoin, is notoriously energy-intensive. Integrating such systems with AI models could escalate operational costs" [M4].

Additional concerns include the quality of AI outputs (e.g., bias and opacity) and the general lack of empirical validation, with few real-world deployments and limited reproducibility [M4, M24, M33, M34, M36, M40, F32]. In practice, BAISET technology stacks provide a trust backbone but often struggle to meet production-level requirements in scalability, efficiency, privacy, and cost (see Fig. 32).

In our RQ4 dataset, energy consumption and sustainability are explicitly mentioned as future work, while many other papers call for efficiency-oriented solutions, such as efficient consensus mechanisms, Layer-2 protocols, and off-chain computation, implicitly targeting reduced energy use and resource demands. Specifically, [M2] emphasizes "energy efficiency" in blockchain-AI integration, while [F32] highlights "sustainable healthcare delivery" as a future research priority. Several studies [M1, M2, M4, M5, M29, B4, F9] propose concrete approaches, including Layer-2 scaling, lightweight consensus protocols, and off-chain training or storage, all designed to reduce computational requirements, improve throughput, and enhance energy efficiency.

Only two articles (3.8%) [M2, F12] explicitly cite energy or sustainability as future directions. However, nearly half of the studies (24 studies, 45.3%) emphasize efficiency-driven methods more broadly. These include lightweight AI models, Layer-2 scaling techniques, off-chain computation, and mobility-based designs that reduce resource overhead while maintaining operational reliability. Smaller clusters (6 studies, 11.3%) explore complementary approaches, such as low-resource consensus mechanisms like Proof-of-Trust (PoT), decentralized storage solutions, and Proof-of-Concept (PoC) on-chain transitions.

Beyond energy and efficiency, several papers highlight opportunities to improve scalability, throughput, and latency without compromising security or decentralization. Proposed strategies include hybrid on-chain/off-chain architectures, adaptive consensus protocols, and optimized storage mechanisms. Some studies further stress the need for standardized benchmarking of blockchain-AI systems to evaluate energy efficiency, resource consumption, and operational performance under real-world conditions.

Future research directions therefore converge on three intertwined goals:

1. Reducing computational and energy overhead through optimized protocols and off-chain solutions.
2. Enhancing system scalability and responsiveness without compromising security.
3. Enabling verifiable, reproducible, and sustainable AI-blockchain deployments across domains such as healthcare, finance, and IoT.

Collectively, these directions underscore the growing recognition that achieving trustworthy AI-Blockchain systems requires balancing performance, sustainability, and practical deployment feasibility.

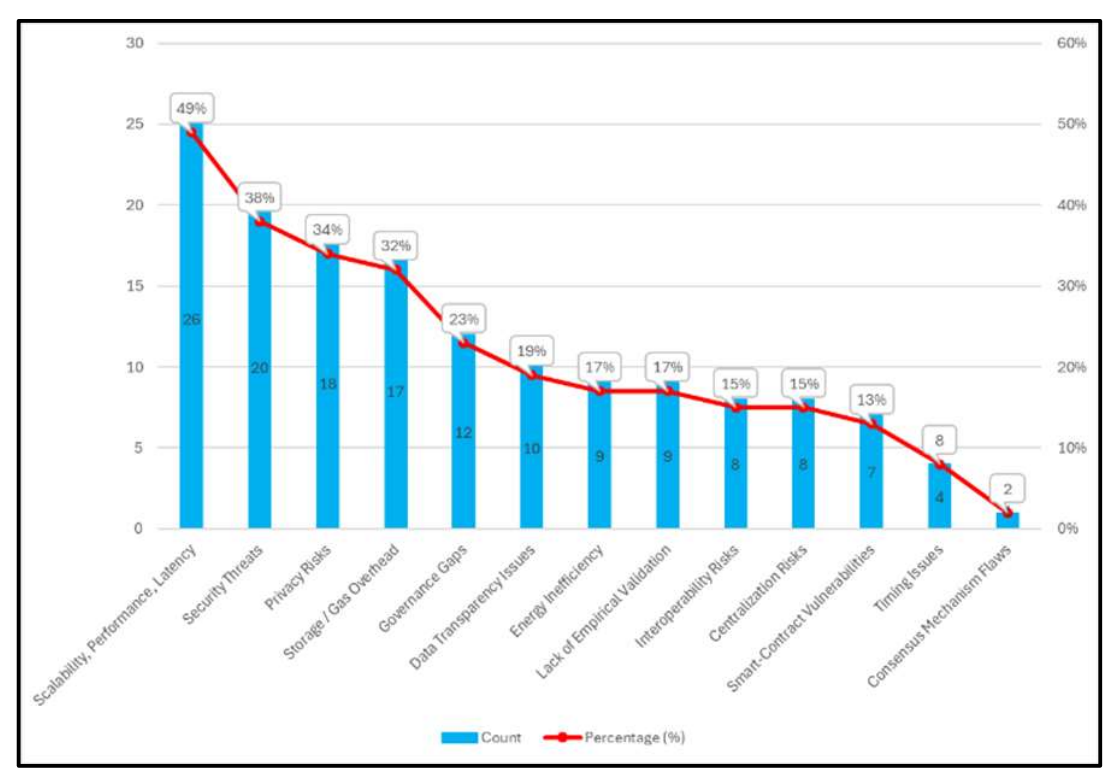

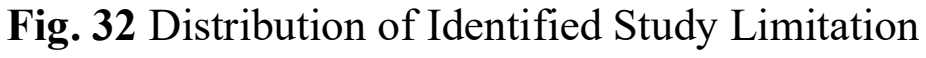

**Fig. 32** Distribution of Identified Study Limitation

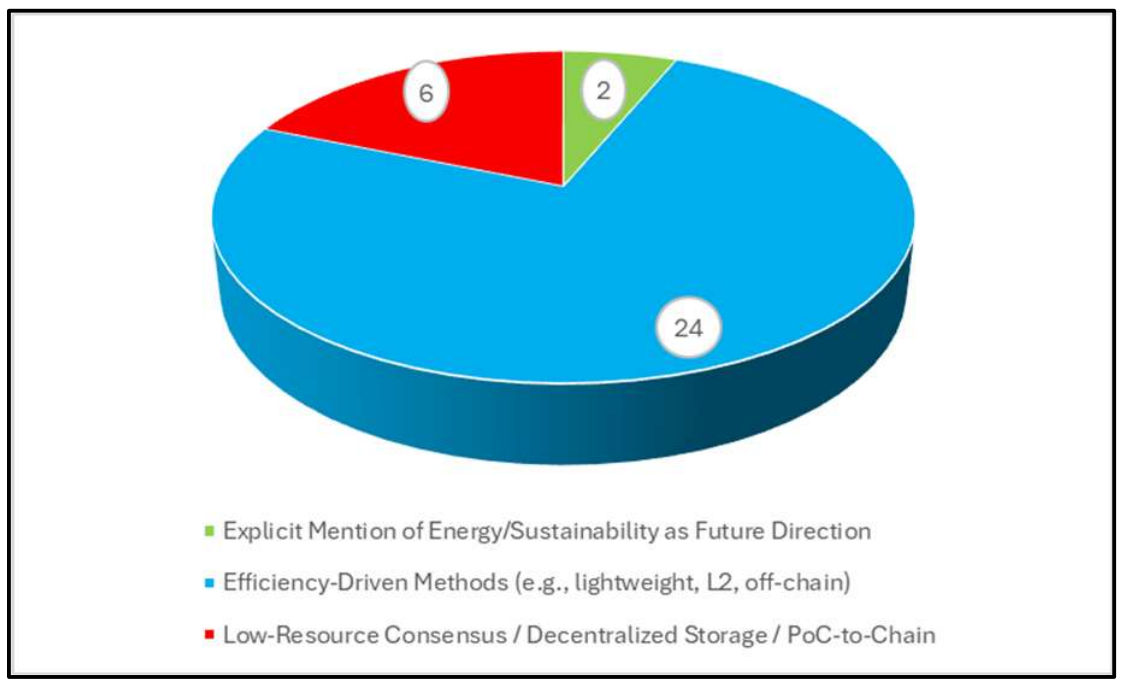

**Fig.33** Distribution of Sustainability and Efficiency

| **RQ4.** What are limitations and recommendations for future research for the use of BAISET? |
| --- |
| Scalability and performance, including latency, throughput, and transaction costs are the primary limitations in integrating blockchain and AI for trustworthiness, reported in 49% of studies. Security and privacy risks (34%-38%), such as adversarial attacks and data vulnerabilities, along with governance challenges like interoperability and regulatory compliance, constitute key secondary constraints. High energy consumption is also a notable barrier (17%), with nearly half of the studies (45%) proposing efficiency-focused solutions, including Layer-2 scaling and off-chain computation, to mitigate this issue. |

## 5. Implications, Roadmap and Future Research Direction

### 5.1 Key Implications

**1. BAISET Research Maturation:** From 2017 to 2025, research on blockchain for AI-supported software evolution (BAISET) has evolved from foundational conceptual frameworks to more applied, domain-specific studies. Regulatory drivers, including the EU AI Act, have accelerated forward-cited research, signaling a shift toward empirical validation and deployable trust mechanisms.

**2. Methodological Diversity and Emerging Trust Dimensions:** Early studies prioritized data integrity and security, while recent work (2020-2025) integrates empirical evaluations, artifact development, and lifecycle-level governance. Beyond traditional trust metrics, research increasingly addresses explainability, auditability, and energy efficiency, reflecting a more holistic understanding of trustworthiness.

**3. Operationalizing Trust in Practice:** Blockchain enables transparency, provenance tracking, and verifiable governance across AI-assisted software evolution. Practitioners should embed trust as an auditable lifecycle artifact, integrating compliance monitoring, update tracking, and data-flow governance to align with regulatory and organizational requirements.

**4. Balancing Trust, Performance, and Sustainability:** Deploying blockchain-enhanced AI introduces trade-offs in latency, throughput, and energy consumption. Solutions such as small-batch synchronization, lightweight consensus algorithms, off-chain computation, and Layer-2 scaling can maintain trust while improving efficiency and sustainability. Sustainability metrics should be evaluated alongside accuracy when assessing system design.

**5. BAISET as a Practical Framework:** BAISET provides a structured approach to operationalizing trustworthy AI: combining blockchain-enabled transparency, governance, and energy-aware design to support scalable, verifiable, and sustainable deployments (Fig.34). Future research should focus on lifecycle-integrated trust mechanisms, standardized benchmarking, and energy-efficient architectures to bridge the gap between conceptual frameworks and real-world adoption.

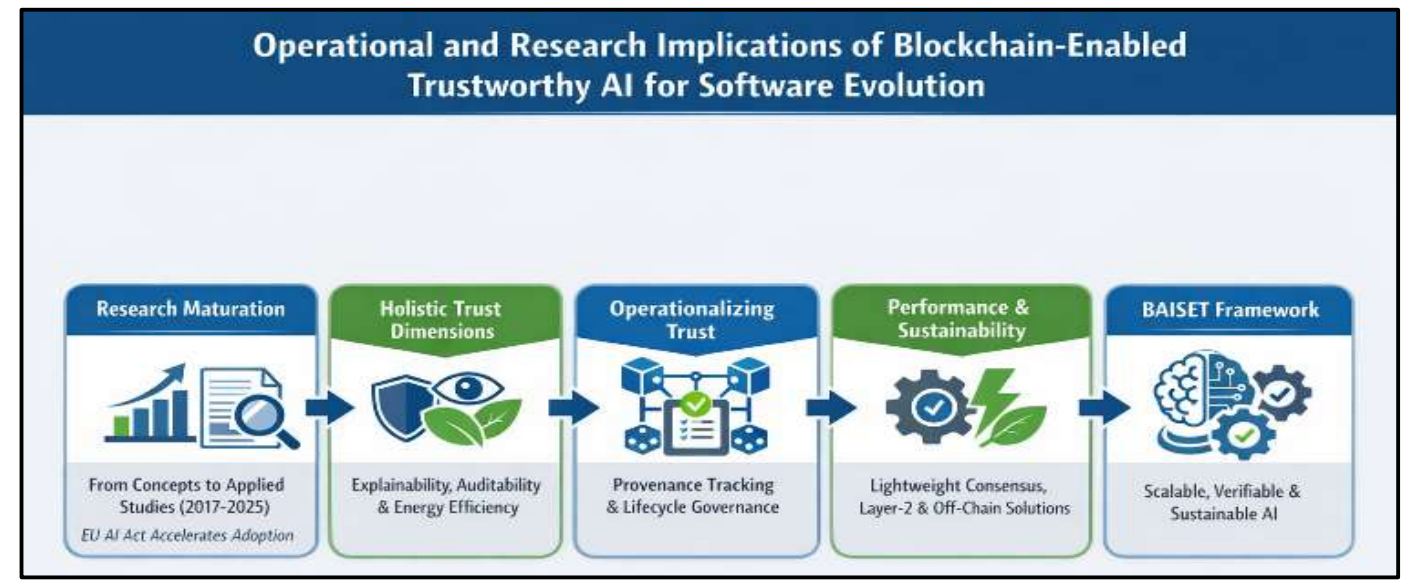

**Fig. 34** Operational and research implications of blockchain-enabled trustworthy AI for software evolution

## 5.3 Roadmap and Future Direction

The following section introduces key short-, medium-, and long-term directions for research and practice in blockchain-enabled trustworthy AI, highlighting scalability, efficiency, governance, and empirical validation as priorities for sustainable AI-Blockchain systems.

The deep learning community is entering a post-scaling era, where efficiency becomes a central design objective. Evidence shows that far more energy-efficient intelligence is possible, the human brain, with roughly 86 billion neurons and trillions of synapses, operates on just tens of watts orders of magnitude less than the power required by large language models on GPUs. Sustainable AI thus represents both a critical challenge and a strategic research opportunity. Achieving comparable or superior performance with dramatically lower energy demands requires rethinking model architectures, learning algorithms, and computation strategies, including dynamic/adaptive computation, quantization, knowledge distillation, and alternative paradigms. Research into non-traditional hardware, such as analog and photonic platforms, further promises substantial reductions in energy consumption. The impact is profound, efficient AI systems reduce environmental footprints, lower barriers for academic and small-scale innovators, enable on-device and decentralized deployment, and support privacy-preserving applications, paving the way for a more accessible, responsible, and sustainable AI ecosystem.

A key limitation of current AI is its lack of interpretability. Deep Neural Networks often function as "black boxes," providing little insight into how predictions are made. This opacity limits their use in high-stakes domains, where understanding, validating, and acting on predictions are critical. Existing interpretable methods remain brittle, unstable, or poorly aligned with true model reasoning, highlighting the urgent need for robust, human-aligned interpretability approaches to ensure safe, accountable, and trustworthy AI.

**Short-Term (1-2 years): Scaling and Standardization**

**Scalability and Efficiency:** Implement Layer-2 scaling, lightweight or alternative consensus mechanisms, hybrid on-chain/off-chain designs, and reduced storage overhead to mitigate latency, throughput, and cost issues [M1, M2, M4, M25, M28, M29, F9].

- **Empirical Validation:** Initiate small-scale deployments, prototype testing, and reproducible evaluation pipelines to establish baseline benchmarks and assess operational feasibility [M4, M14, M24].
- **Privacy and Security:** Begin integrating privacy-preserving frameworks and secure federated learning into pilot systems, focusing on mitigating adversarial attacks and smart-contract vulnerabilities [M1, M4, M7, M21, M30, F24].

**Medium-Term (2-4 years): Energy-Efficiency, Evaluation and Governance**

- **Energy Efficiency and Sustainability:** Deploy off-chain computation, resource-aware architectures, and energy-efficient consensus mechanisms to reduce operational costs and carbon footprint [M2, M5, M9, M29, F8, F18, F32].
- **Governance and Interoperability:** Develop standardized governance frameworks, cross-chain interoperability mechanisms, and regulatory-aligned trust models to ensure ethical and compliant AI-Blockchain operations [M2, M7, M29, M33, M39, B13, F19, F21].
- **Broader Evaluation:** Expand empirical validation to include longitudinal studies and comparative evaluations across application domains. Introduce standardized datasets and benchmark tasks for BAISET systems [M34, M36, M40, M44].

**Long-Term (4-6+ years): Mature and Sustainable BAISET Ecosystems**

- **Integrated Trustworthy AI Systems:** Achieve fully operational AI-Blockchain infrastructures with end-to-end provenance, transparency, auditability, and lifecycle governance.

- **Energy-Conscious, Scalable Deployment:** Systems operate at production scale with minimal environmental impact, maintaining trust and compliance while supporting high-throughput workloads.
- **Evidence-Based Best Practices:** Establish industry-wide benchmarks, regulatory-aligned methodologies, and validated frameworks to facilitate widespread adoption of BAISET solutions across diverse sectors.

**Summary:** The timeline illustrates a progression from foundational architectural and empirical improvements, through scalability, standardization, and governance, to fully integrated, sustainable, and regulation-aligned BAISET systems. This phased approach provides a roadmap for translating current conceptual and prototype-level research into mature, deployable, and trustworthy AI-Blockchain ecosystems.

## 6. Threats to Validity

This SLR follows the guidelines of Kitchenham and Charters [T01, T02] and the framework of Runeson et al. [T03], covering internal, construct, conclusion, and external validity. While designed to ensure reliability and minimize bias, several potential threats remain. The protocol was refined through pilot tests and continuous author discussions to maintain systematic and replicable search, selection, and analysis.

**Construct Validity:** Construct validity concerns whether the methodology measures what it intends [T03-T05]. To strengthen it, we used:

- Primary search: queries across eight peer-reviewed libraries.
- Secondary search: forward and backward snowballing of highly cited, related works.
- Supplementary repositories: sources such as Academia.edu to reduce publication bias.

Search strings were tailored to database syntax and temporal coverage (2017-2025). Pilot testing and continuous expert discussions ensure consistent interpretation of trustworthiness dimensions (security, transparency, privacy, reliability, and explainability). PRISMA tracking further mitigates construct threats.

**Conclusion Validity:** Conclusion validity assesses whether findings logically follow from data [T01, T03]. Data extraction was guided by the PICOC framework and mapped to RQ1-RQ4 across 40 fields per paper. Quantitative counts were triangulated with qualitative analysis. While some studies lacked detailed reporting, aggregation via counts, proportional measures, and citation mapping ensures traceable conclusions. All data and logs were retained for replication.

**External Validity:** External validity examines generalizability [T03, T05]. The review focused on peer-reviewed, English-language studies from 2017-2025, covering AI and blockchain in software engineering. Short papers, vision papers, and non-primary reports were excluded to preserve depth, which may omit gray literature, non-English studies, and industrial validations. Future work should expand to these sources and empirical benchmarks.

**Internal Validity:** Internal validity concerns bias in observed patterns [T03, T05]. The first author designed the initial protocol, validated with co-authors and experts. Pilot testing and repeated calibration during quality assessment (QA1-QA8) ensured consistent selection and coding across 96 studies (44 primary, 20 backward, 32 forward), minimizing judgmental bias.

## 7. Conclusions

Our synthesis reveals that: (1) trustworthiness in AI-supported software engineering is mostly addressed through security, transparency, and general trust, while explainability, verifiability, and auditability remain underexplored (RQ1b); (2) blockchain primarily functions as a foundational trust layer most often via smart contracts whereas LLM-based approaches are still emerging (RQ3); and (3) the field remains methodologically immature, dominated by conceptual or survey-based studies, with no deployed prototypes and limited transparency in datasets and benchmarks, restricting reproducibility and systematic evaluation (RQ3).

Future research should focus on scalable, efficient architectures, empirical validation through prototypes and benchmark pipelines, privacy- and security-preserving designs, and governance and interoperability mechanisms aligned with standards and regulations. Pursuing these directions will help advance BAISET from conceptual frameworks toward practical, deployable, and evidence-based solutions.

## Author Contributions

Mohammad Naserameri conceived the main research idea and designed the overall structure of the study. He carried out the systematic literature search, study selection, data extraction, and synthesis of the reviewed studies. Mohammad Naserameri drafted the original manuscript and implemented all revisions. The academic supervisor contributed to the conceptual refinement of the research scope, provided methodological guidance throughout the study, critically reviewed the manuscript for scientific rigor and clarity, and supervised the research process. All authors read and approved the final version of the manuscript.

## Statements and Declarations

**Competing Interests.** The authors declare that they have no competing interests.
**Funding.** This research received no external funding.
**Ethical Approval.** This study is based exclusively on previously published literature. No human participants or animals were involved, and therefore ethical approval was not required.
**Data Availability.** All data analyzed during this study are included in this published article and its referenced sources.

The dataset generated and analyzed during this study is publicly available in the Zenodo repository at: [https://doi.org/10.5281/zenodo.18405620].

## Appendix A.

***List of Primary*** *Studies (Labeled with "M"),* ***Backward Snowballing Studies*** *(Labeled with "B")* ***and List of Forward Snowballing*** *Studies (Labeled with "F")*